\documentstyle[prb,aps,multicol,ifthen,epsfig]{revtex}
\newcommand{\wide}[2]{
\end{multicols}
\widetext
\noindent
\ifthenelse{\equal{#1}{t}}
{}
{
\raisebox{0.1in}[0in][0.02in]{$\rule{3.575in}{0.002in}
\rule{0.002in}{0.08in}$}
}
#2
\ifthenelse{\equal{#1}{b}}
{}
{
{\raisebox{-0.1in}[0in][0.02in]
{\hspace{3.575in}$\rule{0.002in}{0.08in}
\rule[0.08in]{3.575in}{0.002in}$}
}
}
\begin{multicols}{2}
\noindent
}

\tighten  %
\begin{document}

\title{`Tight Binding' methods in quantum transport through molecules and small
devices: From the coherent to the decoherent description.}
\author{Horacio M. Pastawski$^{1}$ and Ernesto Medina$^{2}$}
\address{$^{1}$ Facultad de Matem\'{a}tica, Astronom\'\i a y F\'\i sica,
Universidad Nacional de C\'{o}rdoba, Ciudad Univeritaria, 5000 C\'{o}rdoba,
Argentina}
\address{$^{2}$ Laboratorio de F\'\i sica Estad\'\i stica de Sistemas 
Desordenados, Centro de F\'\i sica, Instituto Venezolano de
Investigaciones Cient\'\i ficas, Caracas, Venezuela}
\maketitle

\begin{abstract}
We discuss the steady-state electronic transport in solid-state and
molecular devices in the quantum regime. The decimation technique
allows a comprehensive description of the electronic structure. Such a
method is used, in conjunction with the generalizations of Landauer's
tunneling formalism,  to describe a wide range of transport regimes. We
analize mesoscopic and semiclassical metallic transport, the metal-insulator
transition, and the resonant tunneling regime. The effects of decoherence on
transport is discussed in terms of the D'Amato-Pastawski model. A brief presentation
of the time dependent phenomena is also included.	
\\

Se discute el transporte, en el estado estacionario, en dispositivos moleculares y de estado s\'{o}lido en el r\'{e}gimen cu\'{a}ntico. La t\'{e}cnica de decimaci\'{o}n permite una descripci\'{o}n completa de la estructura electr\'{o}nica. Tal m\'{e}todo, en conjunto con la generalizaci\'{o}n del formalismo de Landauer, puede ser usado para describir un rango amplio de reg\'{i}menes de transporte. Se analiza el transporte mesosc\'{o}pico y semicl\'{a}sico en el r\'{e}gimen met\'{a}lico, la transici\'{o}n metal-aislante, y el r\'{e}gimen de t\'{u}nel resonante. Los efectos de decoherencia en el transporte son tratados en t\'{e}rminos del modelo de D'Amato-Pastawski. Finalmente se incluye una breve presentaci\'{o}n de fen\'{o}menos dependientes del tiempo.

\textit{Descriptores:} Coherencia cu\'{a}ntica; funciones de Green; tunelamiento resonante

\end{abstract}


\begin{multicols}{2} %
\section{Introduction}

\begin{figure}[tbp]
\narrowtext
\centering \leavevmode
\center{\epsfig{file=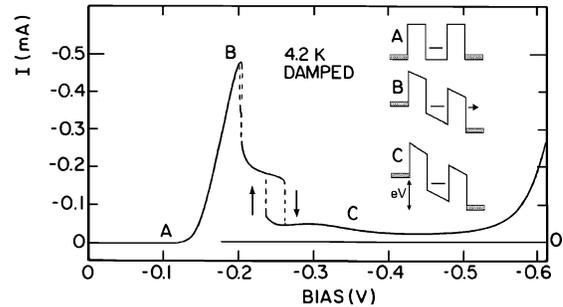,width=7.5cm}} 
\vspace{0.5cm}
\caption{Scheme of a DBRTD and experimental current-voltage characteristic. 
Results adapted from Ref. 2. 
The points A, B and C in the curve corresponds to the 
potential profile shown in the inset.}
\label{fig_exp-barrier}
\end{figure}

With the advent of progressive miniaturization of electronic devices,
it becomes necessary to describe the transport properties of small
systems within a fully quantum mechanical framework. As a representative
example, we just recall the double barrier resonant tunneling
device\cite{z--Esaki} (DBRTD), whose current-voltage ($\mathsf{I-V}$)
curve presents a well defined peak as shown in
Fig.\ref{fig_exp-barrier} adapted from data in reference
\cite{z--T+e-phexpt}. This behavior, in great departure from the usual 
monotonic features shown by macroscopic samples, can be attributed to
quantum interference phenomena, which arises from the coherent
dynamics of the carriers.

Let us characterize a ``device'', in a rough sense, as some specified
region of material where the carriers spend a substantial part of the
time. It might be characterized by a longitudinal length $L_{x}$ and a
transverse cross section $L_{y}\times L_{z}=M\,a^{2},$ where $a$ is an
atomic length scale. In order to manifest its wave nature, an
electronic excitation must propagate quantum mechanically (i.e. phase
coherently) between these boundaries. Typical interference phenomena
occurs when any of these lengths is comparable with the deBroglie 
wavelength $\lambda_{\varepsilon}.$ A condition for this quantum coherence
is the weak coupling with the environmental degrees of freedom within
the relevant time scales. This condition can often be achieved by
shrinking the device length scales and/or lowering the temperatures
until the minimum energy excitation $\Delta$ becomes large compared to
the available thermal energy $k_{B}T$. In the experiment of
Fig.\ref{fig_exp-barrier} the applied voltage is the control
parameter for the wave length of the electrons responsible for the
transport. Therefore, the peaked ${\mathsf I-V}$ curve of the DBRTD can be viewed
much as resonant peaks in a Fabry-Perrot interferometer.

A great variety of solid-state devices developed in the last decade
\cite{z--mesoscopic-books} satisfy the general conditions for the quantum
manifestations described above. More recently, the growing interest
\cite{z--T-Molecular/SciAm} in understanding the conductance between two
electrodes connected through a bridging molecule lies naturally in this
category. In fact, discreteness of the molecular electronic energy levels is a
clear manifestation of the quantum coherence of the electronic states and one
is left with the question of how do these states transport charge
\cite{z--Mikkelsen-Ratner}. The related issue of the propagation of a charge
density excitation has been tackled by physical-chemists who, for a long
time, have been dealing with problems as diverse as conducting polymers
\cite{z--SSH}, charge oscillation in the intermediate valence compounds
\cite{z--Taube} (e.g. the Creutz-Taube ion (NH$_{3}$)$_{5}$Ru-pyz-Ru-(NH$_{3}%
$)$_{5}^{\,\,\,\,\,+5}$ where pyz stands for pyrazine) and electron transfer
in photosynthetic systems \cite{z--Beratan}.

The traditional methods used to describe transport are not necessarily
appropriate for mesoscopic and molecular systems. The semiclassical
Boltzmann equation, the Kubo formalism and other traditional
techniques were specially devised to describe bulk transport, where
the thermodynamic limit is guaranteed and the linear response regime
is the relevant one. Hence, in order to deal with transport in finite
size ``samples'' with non trivial geometrical constrains, and often in
presence of non-perturbative fields, a new perspective had to be
adopted. It is clear that a general Schr\"{o}dinger equation must
provide the complete description. However, even in the simplest
independent particle approximation, it seems difficult to include the
complex boundary conditions imposed by the electrodes. A way out is to
treat transport of carriers through the ``sample '' as a scattering
problem from one electrode to the other. Therefore, the transmittance
$T$ \ undertakes a most relevant role. This is essentially the
conceptual framework on which the Landauer's
description\cite{z--Landauer} of transport is based. It involves a
simple but conceptually new approach that inspired most of the
advances in solid-state devices in recent years and is now clearing
the way towards molecular electronics.

In this article, we attempt to present a self-contained operational
and conceptual manual based on the lectures given by the authors at
the 2nd Workshop on Mesoscopic Systems. We will attempt to summarize
the methods used to evaluate and interpret the electronic structure of
finite systems, as well as how they are extended to describe the
coherent transport when they are connected to electrodes. While 
the effects of electron-electron
interaction can become of utmost relevance in small
systems\cite{z--Kastner}\cite{z--Devoret}, for didactic
reasons we will not extend on this blooming field. Let's only mention that
their understanding\cite{z--review-DOT} is built upon the methods and ideas we describe in this review.

The material is presented as follows: Section II is devoted to
Landauer's ideas, and represents the backbone over which the formalism
is built. Section III introduces the decimation method to calculate
transmittances and electronic structure of devices and molecules.
Section IV shows the connection between electronic-structure and
transport obtained by Fisher and Lee. Their formula is derived and
discussed there. Section V presents an overview of the known results
of transport which can help the beginner to see the machinery in
action. Section VI is devoted to the introduction of decoherence into
the formalism. Section VII presents a brief outline of the
time-dependent problem.

A number of the most simple calculations will be worked out with some
detail, so the unfamiliar reader might acquire a first feel for the
topic by noting the simplicity of the methodology. The most advanced
results will be left merely indicated so that they can be reobtained
by a motivated reader. Many of the examples presented here are
completely original and have not been published before to our
knowledge. Finally, let us stress that we do not attempt to review the
literature exhaustively, but simply to present our personal pathway
through the concepts involved. In particular, the issues of
decoherence and time dependence are the object of current research and
the reader might still create his own trek.

\section{The basic ideas of Landauer}

Rolf Landauer was the first to realize that, besides the ``sample ''
or device one must explicitly incorporate the electrodes or contacts
in the transport description.  He introduced two one-dimensional
wires, left ($L$) and right (${R}$), connecting ``the
sample'' to electron reservoirs. The net current leaving the lead
${L}$ is expressed in terms of the \ number of electrons
available in the leads, their typical velocity and the probability
$T_{{R,L}}$ of transmission through the sample
\begin{eqnarray}
{\mathsf
I}_{L}=e\int[T_{R,L}(\varepsilon)v_{L}\frac{1}{2}N_{L}(\varepsilon)
f_{L}(\varepsilon) \nonumber \\
-T_{L,R}(\varepsilon)v_{R}\frac{1}{2}N_{R}(\varepsilon)
f_{R}(\varepsilon)]\mathrm{d}\varepsilon.
\label{eq_Current}
\end{eqnarray}
The meaning of this equation is obvious: it balances currents. Each
reservoir $i$ emits electrons with an energy availability controlled
by a local distribution function
$f_{i}(\varepsilon)=1/[\exp[(\varepsilon -\mu_{i})/k_{B}T+1]$, where
$\varepsilon$ is the energy and $\mu_{i}=\mu_{o}+\delta\mu _{i}$ is
the chemical potential which is displaced from its equilibrium value
$\mu_{o}$. The density of those ``outgoing'' states is
$\frac{1}{2}N_{i}(\varepsilon)$ (half the total) and\ their velocity
$v_{i}$. The coefficient $T_{{R,L}}$ computes current as positive
provided that the particle can pass through the sample. Taking the $i=L$
reservoir as reference, the first term is then an ``out '' current
while the second is the \ ``in'' contribution. It was essential in
Landauer's reasoning to note that in a propagating channel the density
of states $N_{i}$ is inversely proportional to the corresponding group
velocity:$\,$%
\begin{equation}
N_{i}\equiv \frac{2}{v_{i}h}. 
\label{eq_N-v}%
\end{equation}
This fundamental fact remained unnoticed in the previous discussions
of quantum tunneling\cite{z--Esaki} and it is the key to understand
conductance quantization. Notice that there is no use for the
traditional $\left[ 1-f_{j}(\varepsilon)\right] $ factor to
exclude transitions to occupied final states. In a
scattering formulation, the final ``out'' states are already contained
in the ``in'' states\cite{z--Schiff}.  Although different ``in''
states (e.g. on the left and right leads) could end in the same final
state, unitarity of quantum mechanics assures that outgoing
contributions are orthogonal.

\begin{figure}[tbp]
\narrowtext
\centering \leavevmode
\center{\epsfig{file=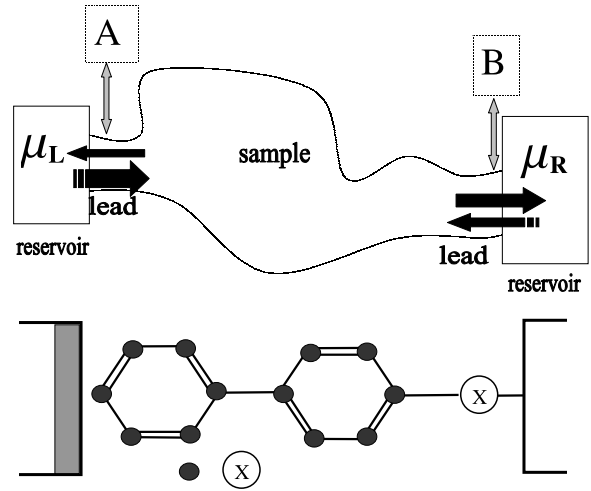,width=8.0cm}} 
\vspace{0.5cm}
\caption{Upper panel: Landauer's representation of a general electronic
transport experiment. Lead at left (L) and right (R) inject and extract
current. Weakly coupled voltage probes A and B at the ``sample'' boundaries
are also represented. Lower Panel: Portion of an actual molecular
sandwich heterostructure between a Au-Ti left electrode and a right Au
electrode. The bridge molecule is 4-4-thioacetilbiphenyl. The residue at right
(X =S, Se, Te) can be substituted.}
\label{fig_reserv+molec}
\end{figure}

To obtain the usual Landauer \textit{two probe} conductance, one
assumes time reversal for the transmittances:
$T_{{R,L}}=T_{{L},{R}}.$ At low temperatures, the
Fermi distribution function can be safely replaced by a step
function. If \textit{linear response} can be invoked ($\delta\mu
_{i}=\mu_{i}-\varepsilon_{F}\ll\varepsilon_{F}$), the integral is
approximated by using the transmittance evaluated at a Fermi
energy. If we do not include the usual factor of 2 due to the spin
degeneracy, one gets
\begin{equation}
{\mathsf I}=\frac{e}{h}T_{{R,L}}(\delta\mu_{{L}}-\delta
\mu_{{R}})\,\,{\rm where\,\,}(\delta\mu_{{L}}-\delta
\mu_{{R}})=e{\mathsf V} 
\label{eq_Landauer-current}
\end{equation}
from which the two-probe conductance can be calculated
\begin{equation}
{\mathsf G}_{{R,L}}=\frac{e^{2}}{h}T_{{R,L}}.
\label{eq_G-landauer}
\end{equation}
This conductance accounts for the ``sample-lead'' system. In a perfect
1-d conductor $T_{{R,L}}=1$, and it remains finite. In fact,
resistance results $h/e^{2}=25.812K\Omega$ per channel, the quantum
value found later in the Quantum Hall Effect. In contrast, the
original Landauer's conductance
\begin{equation}
{\mathsf G}_{{R,L}}^{{\rm f.p}.}=\frac{e^{2}}{h}\frac{T_{{R,L}
}}{1-T_{{R,L}}} 
\label{eq_4p-Landauer}
\end{equation}
is an attempt to extract an intrinsic property of the ``sample''. This
conductance corresponds to a \textit{four probe measurement, }as shown
in Fig. \ref{fig_reserv+molec} (upper panel) where two leads are used to
inject and drain the current respectively, while the other two
additional ``non-invasive'' probes, denoted by A and B, are used to
measure the voltage drop in the ``sample'' neighborhood. They are very
weakly coupled wires or capacitive ``non-invasive'' probes. The
difference,
\begin{equation}
{\mathsf R}^{{\rm contact}}=1/{\mathsf G}_{{R,L}}-1/{\mathsf
G}_{{R,L}}^{{\rm f.p.}}=\frac{h}{e^{2}},
\label{eq_contact-res}
\end{equation}
can be interpreted as a contact resistance associated to the
lead-sample interface.

A more general formulation of Eqs.(\ref{eq_Current}) and (\ref{eq_N-v}) for
a system composed by many channels (e.g. spin and transversal modes) subject to
different boundary conditions results from the application
\cite{z--Büttiker-Kirchhoff} of the Kirchoff law using Landauer's conductances
\begin{equation}
{\mathsf I}_{i}=\frac{e}{h}\sum_{j}\int[T_{j,i}(\varepsilon){f}
_{i}(\varepsilon)-T_{i,j}(\varepsilon){f}_{j}(\varepsilon)]\mathrm{d}\varepsilon. 
\label{eq_current-multilead}
\end{equation}
We do not exclude sites $i=j$ from the sum. Here, the transmission
coefficients may depend on the external parameters such as voltages,
and hence accounts for non-linear response. The non-ohmic ${\mathsf
I-V}$ curve of the DBRTD shown in Fig.\ref{fig_exp-barrier} can indeed
be obtained directly from this formula by summing up over transversal
quantum numbers (channels) in the left and right leads. Also
$\mu_{i}=\mu_{{L}}$ for every outgoing channel in the left lead and
$\mu_{i}=\mu_{{R}}$ for every outgoing channel in the right
lead.

Having understood the basic requirements of a transport theory one
must learn to connect the transmittances $T$ with the electronic
properties of the ``sample'' and electrodes. In the treatment of solid
state devices many researchers adopted the strategy of modeling the
transport through the scattering matrices. However, when one deals
with a molecular system, it becomes clear that specific features of
the electronic structure at the molecular level are relevant. Thus, at
least an approximate description of these properties through
Hamiltonian models is mandatory. Let us remark that in the Landauer
formalism, the ``sample'' is considered a finite system while the
electrodes are considered in their thermodynamic limit with a
continuum spectrum and acting as charge reservoirs. One then needs a
formulation capable of dealing naturally with both situations.

\section{ Electronic Properties through Real Space Renormalization Group 
procedures.}

\subsection{Origin of the tight-binding model}

The first thing we want to point out is that tight binding models, although
naturally associated with {\it linear combination of atomic orbitals} (LCAO) can
also be obtained as a convenient approximation to the Schr\"{o}dinger equation
written in terms of the continuous coordinate $x$
\begin{equation}
-\frac{\hbar^{2}}{2m}\nabla^{2}\psi(x)+U(x)\psi(x)=\varepsilon\psi(x)
\label{eq_Schr-1d}%
\end{equation}

We can discretize this equation obtaining a finite differences approach
\wide{m}{
\begin{equation}
-\frac{\hbar^{2}}{2m}\displaystyle\frac{\displaystyle\frac{\psi(x+\Delta x)-\psi(x)}{\Delta x}%
-\displaystyle\frac{\psi(x)-\psi(x-\Delta x)}{\Delta x}}{\Delta x}+U(x)\psi(x)=\varepsilon
\psi(x) \label{eq_finite-diff}
\end{equation}
}
If we do the identifications
\begin{equation}%
\begin{tabular}
[c]{l}%
$\Delta x=a;\quad x=na;\quad u_{n}=\psi(na);$\\
$E_{n}=U(x_{n});$ $\,\,\,V=\frac{\hbar^{2}}{2ma^{2}};$
\end{tabular}
\label{eq_discr/cont}
\end{equation}
we finally obtain:
\begin{equation}
(\varepsilon-E_{n})u_{n}-Vu_{n+1}-Vu_{n-1}=0. \label{eq_TB-difference}%
\end{equation}
Therefore, one is left with a discrete equation where the interaction
is provided by the \textit{kinetic energy} terms $V$ which is usually
short ranged. The \textit{local potential energy }term is given by
$E_{n}$ and, in the LCAO description, it can be identified with the
energies of atomic orbitals.

The above relation can be used in a ``reversible'' way since there are
situations where calculations in the continuum are simpler, as occurs
in the semiclassical limit when the WKB approsimation substantially simplify the
problem.

Indeed, when the discretization is done at a scale much smaller than
the typical atomic scale $\Delta x\ll a_{0}$ one can reach the
molecular scale of the LCAO by a progressive use of the procedure that
is discussed in the following paragraph. For the moment, we consider
the tight binding Hamiltonian:
\wide{m}{
\begin{equation}
\hat{\mathcal{H}}=\sum\limits_{n}E_{n}\left| n\right\rangle \left\langle
n\right| +V_{n,n+1}\left| n\right\rangle \left\langle n+1\right|
+V_{n+1,n}\left| n+1\right\rangle \left\langle n\right| ,
\label{eq_1d-TB}%
\end{equation}
with$\;\left\langle n\right.  \left| \psi\right\rangle =u_{n}$ ,
$E_{n}\equiv E_{o}$ and $V_{n,n+1}=-V,$ from which the Schr\"{o}dinger
Eq.(\ref{eq_1d-TB}) takes the matrix form
\begin{equation}
\varepsilon\left(
\begin{array}
[c]{c}%
\vdots\\
u_{n-1}\\
u_{n\,\,\,\,\,\,\,}\\
u_{n+1}\\
\vdots
\end{array}
\right)  -\left (
\begin{array}
[c]{ccccc}
\ddots &  &  &  & \\
&  E_{n-1} & V_{n-1,n} & 0 & \\
& V_{n,n-1} & E_{n} & V_{n,n+1} & \\
& 0 & V_{n+1,n} & E_{n+1} & \ddots\\
&  &  &  & \ddots
\end{array}
\right )  \left(
\begin{array}
[c]{c}%
\vdots\\
u_{n-1}\\
u_{n\,\,\,}\\
u_{n+1}\\
\vdots
\end{array}
\right)
  =\left(\varepsilon\mathbf{I}-\mathbf{H}\right)  \overrightarrow
{u}=\overrightarrow{0}.
\label{eq_1d-(E-H)u}\\
\end{equation}
}
When the site energies are taken from a random distribution in the range [$-W/2,W/2$] the Hamiltonian is referred to as the Anderson
model and it is the standard model to represent disordered systems.

In a general situation of a LCAO problem, $E_{n}$ corresponds to any
of the specific atomic orbitals used for each atom and the
interactions parameters might not be restricted to first neighbors
complicating the topology of the interaction network.

\subsection{Effective Hamiltonians and Green's functions}

Let's see how to construct a progressive solution of this
problem\cite{z--Levstein-decim} using the ideas of the real space
renormalization group \cite{z--Löwdin,z--Kadanoff}.

\begin{figure}[tbp]
\narrowtext
\centering \leavevmode
\center{\epsfig{file=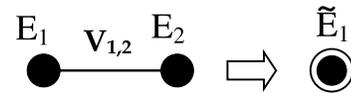,width=4.5cm}} 
\vspace{0.5cm}
\caption{Construction of the self-energy correction.}
\label{fig_2atom+decim}
\end{figure}

The simplest case is the two site problem
\begin{equation}
\varepsilon\left(
\begin{array}
[c]{c} 
u_{1}\\ u_{2}
\end{array}
\right)  -\left (
\begin{array}
[c]{cc}%
E_{1} & V_{1,2}\\
V_{2,1} & E_{2}%
\end{array}
\right )\left(
\begin{array}
[c]{c}%
u_{1}\\
u_{2}%
\end{array}
\right)  =0, \label{eq_2x2 _(E-H)u}%
\end{equation}
whose energy spectrum is obtained from the secular equation
\begin{equation}
\det\left|  \varepsilon\mathbf{I}-\mathbf{H}\right|  =0, \label{eq_secular}%
\end{equation}
with eigenvalues
\begin{equation}
E_{\pm}=\left(  \frac{E_{1}+E_{2}}{2}\right)  \pm\sqrt{\left(  \frac
{E_{1}-E_{2}}{2}\right)  ^{2}-V_{1,2}V_{2,1}}.
\end{equation}
An alternative procedure is to write the linear equation explicitly
\begin{equation}
\begin{array}
[c]{c}
a)\,\,\,\,\,E_{1}u_{1}+V_{1,2}u_{2}=\varepsilon u_{1},\\
b)\,\,\,\,\,V_{2,1}u_{1}+E_{2}u_{2}=\varepsilon u_{2}.
\end{array}
\end{equation}
From $b)$ $u_{2}=V_{2,1}\frac{1}{\varepsilon-E_{2}}u_{1}$, and substituting
in $a)$:
\begin{equation}
\left(  E_{1}+V_{1,2}\frac{1}{\varepsilon-E_{2}}V_{2,1}\right)
u_{1}=\varepsilon u_{1}. \label{eq_eff-(E-H)u-2x2}%
\end{equation}
The eigenvalue is obtained from the condition
\begin{equation}
\begin{array}
[c]{ccc}
\varepsilon-[E_{1}+ & \underbrace{V_{1,2}\frac{1}{\varepsilon-E_{2}}V_{2,1}}] 
& =0\\
& \Delta_{1}(\varepsilon) &
\end{array}
\label{eq_EFFsecular2x2}
\end{equation}
The reader can check that \emph{both exact eigenenergies}, can be
obtained from this equation. The second term in the parenthesis has a
clear physical meaning if it is identified with an ``effective
potential'' $\Delta_{1} (\varepsilon)=\frac{\left| V_{1,2}\right|
^{2}}{\varepsilon-E_{2}}$ which corrects the non-interacting energy
$E_{1}$ of the site as
\begin{equation}
\widetilde{E}_{1}=E_{1}+\Delta_{1}(\varepsilon). 
\label{eq_renormE}
\end{equation}
For an asymmetric case, $\left|E_{1}-E_{2}\right| >\left|  V_{1,2}\right|,$
one can obtain a good approximation to the corrected site energies, by
doing the evaluation at the old eigenvalue:
\begin{equation}
\widetilde{E}_{1}\simeq E_{1}+\Delta_{1}(E_{1}), 
\label{eq_approxi}
\end{equation}
which becomes equivalent to the second order Rayleigh-Schr\"{o}dinger
perturbation theory. That is, as shown in
Fig.\ref{fig_2atom+decim}, we have gone from the two orbital
problem to a single effective orbital with a ``dressed energy ''.

\subsection{The Decimation Method}

\begin{figure}[tbp]
\narrowtext
\centering \leavevmode
\center{\epsfig{file=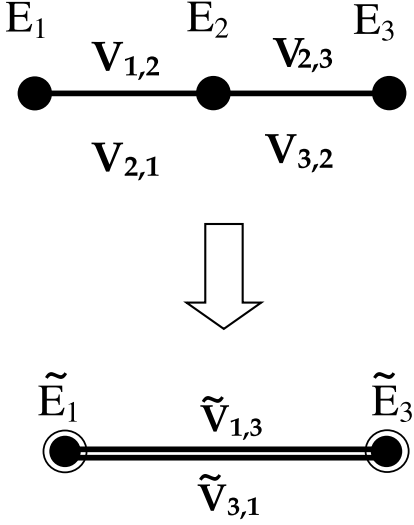,width=6.0cm}} 
\vspace{0.5cm}
\caption{Construction of an effective Hamiltonian through decimation.}
\label{fig_3atom+2eff}
\end{figure}

The preceding paragraphs have introduced the basic tools by noting that the presence of other sites has the effect of shifting or splitting the energies of the site we are looking
at. Let us see another example, a Hamiltonian with three sites:%

\begin{equation}
\left(
\begin{array}
[c]{ccc}
E_{1} & V_{1,2} & V_{1,3}\\
V_{2,1} & E_{2} & V_{2,3}\\
V_{3,1} & V_{3,2} & E_{3}
\end{array}
\right)  \left(
\begin{array}
[c]{c}
u_{1}\\
u_{2}\\
u_{3}
\end{array}
\right)  =\varepsilon\left(
\begin{array}
[c]{c}
u_{1}\\
u_{2}\\
u_{3},
\end{array}
\right)  
\label{eq_3x3(E-H)u}
\end{equation}
where we have included second neighbor overlaps. The idea is to reduce this set of
three linear equations to a smaller problem as shown in Fig.\ref{fig_3atom+2eff}.
There are many situations where the physics and chemistry indicates that this is a
physically meaningful model. We obtain the value of $u_{2}$ from the
second row above
\begin{equation}
u_{2}=\frac{V_{2,1}}{\varepsilon-E_{2}}u_{1}+\frac{V_{2,3}}{\varepsilon-
E_{2}}u_{3}, 
\label{eq_3x3_u2}
\end{equation}
and introduce it into the first and third row to obtain two coupled
non-linear equations:
\begin{equation}
\left (
\begin{array}
[c]{cc}
\widetilde{E}_{1} & \widetilde{V}_{1,3}\\
\widetilde{V}_{3,1} & \widetilde{E}_{3}
\end{array}
\right )  \left(
\begin{array}
[c]{c}
u_{1}\\
u_{3}
\end{array}
\right)  =\varepsilon\left(
\begin{array}
[c]{c}
u_{1}\\
u_{3}
\end{array}
\right)  , \label{eq_3x3-->2x2}
\end{equation}
where
\begin{equation}
\begin{array}
[c]{c}
\widetilde{E}_{1}(\varepsilon)=E_{1}+V_{1,2}\displaystyle\frac{1}{\varepsilon-E_{2}}V_{2,1},\\
\widetilde{E}_{3}(\varepsilon)=E_{3}+V_{3,2}\displaystyle\frac{1}{\varepsilon-E_{2}}V_{2,3},\\
\widetilde{V}_{1,3}=V_{1,3}+V_{1,2}\displaystyle\frac{1}{\varepsilon-E_{2}}V_{2,3},
\end{array}
\label{eq_3x3-->2x2paramet}
\end{equation}
and the notation is self-explanatory.

Therefore, the procedure of elimination of variables is very general and can
exhaust the degrees of freedom of the finite system, providing a systematic
way to reduce the dimension of the Hamiltonian. This is done at the cost of
transforming the linear equation into a non-linear one which, however, can
often be linearized in the region of interest as in Eq. (\ref{eq_approxi}).
The general recipe valid for a string of interacting atoms requires the
calculation of:

\begin{itemize}
\item $\Delta_{1(n)}^{+},$ the energy correction to the $1^{\rm st}$ atom 
when sites until atom $n$, to the right (+) inclusive, are eliminated.

\item $\Delta_{n+1}^{-}(\varepsilon),$ the energy correction to atom
($n+1$) when all the atoms at the left (-), except the $1^{\rm st}$,
have been eliminated.

\item $\widetilde{V}_{1,n+1},$ the effective interaction among layers.
\end{itemize}

We resort to the recursion formulas:
\begin{eqnarray}
\Delta_{1(n)}^{+}(\varepsilon)  & = &\Delta_{1(n-1)}^{+}+\widetilde{V}_{1,n}%
\frac{1}{\varepsilon-E_{n}-\Delta_{n}^{-}}\widetilde{V}_{n,1}
\label{eq_DeltaLEFT}\\
\Delta_{n+1}^{-}(\varepsilon)  &  = &V_{n+1,n}\frac{1}{\varepsilon-E_{n}%
-\Delta_{n}^{-}}V_{n,n+1}\label{eq_DeltaRIGHT}\\
\widetilde{V}_{1,n+1}  &  = &\widetilde{V}_{1,n}\frac{1}{\varepsilon-E_{n}-\Delta
_{n}^{-}}V_{n,n+1}+V_{1,n+1} \label{eq_Veff-iterattive}
\end{eqnarray}
Notice that we have not restricted the iterations to nearest
neighbors. From the point of view of the perturbation theory, it can
be checked that the decimation procedure is equivalent to an exact
summation of all the perturbation orders of the Wigner-Brillouin
series and hence equivalent to the inclusion of all Feynman paths that
start at layer $0$ and end at layer $N+1$. This will be best
viewed in the language of the Green's function described in the next
subsection.

\subsection{Effective Hamiltonians and Green's functions}

The Green's function provides an alternative framework to discuss the
solutions of the Schr\"{o}dinger equation. Besides the intuitive structure of the
perturbative calculations in terms of Feynman diagrams\cite{z--Mattuck}, they
have the additional advantage of a clear connection to transport properties,
and finally a special role in the Quantum Field Theory which allows a
definitive treatment of the many-body problems.

The retarded (advanced) Green's functions matrix is defined as a function of
the complex variable $(\varepsilon\pm \mathrm{i}\eta)$ in the upper (lower)
complex plane.
\begin{equation}
{\mathbf G}^{R(A)}(\varepsilon)=[(\varepsilon\pm{\rm i}
\eta)\mathbf{I}-\mathbf{H}]^{-1}. 
\label{eq_G-matrix}
\end{equation}
Let's evaluate a diagonal element:
\begin{eqnarray}
G_{1,1}^{R}(\varepsilon+{\rm i}\eta)& = &\frac{1}{
\begin{array}
[c]{cc}
\varepsilon+{\rm i}\eta-E_{1}- & \underbrace{V_{1,2}\frac{1}
{\varepsilon+{\rm i}\eta-E_{2}}V_{1,2}}\\
& \Sigma_{1}^{R}=\Delta_{1}-{\rm i}\Gamma_{1}%
\end{array}
}\nonumber \\
& = &\frac{1}{\varepsilon+{\rm i}\eta-\widetilde{E}_{1}}.\label{eq_Green2x2}
\end{eqnarray}
Therefore the renormalized site energy is as before
\begin{equation}
\begin{array}
[c]{cc}
\widetilde{E}_{1}=
E_{1}+ & \underbrace{V_{1,2}\frac{1}{\varepsilon+{\rm i}\eta-E_{2}}V_{2,1}}\\
& \Sigma_{1}^{R}=\Delta_{1}-{\rm i}\Gamma_{1}
\end{array}
\label{eq_new-energy2x2}
\end{equation}
The local densities of states represents the weight of the exact
eigenenergies on the old states,
which we can check for our simple two state model, is:
\begin{eqnarray}
N_{1}(\varepsilon)  
&  = & \left|  u_{1,+}\right|  ^{2}\delta(\varepsilon-E_{+})+\left|
u_{1,-}\right|  ^{2}\delta(\varepsilon-E_{-}) \label{eq_LDOS2}\\
&  = &\lim_{\eta\rightarrow0}\left\{  \left|  u_{1,+}\right|  ^{2}\frac{1}{\pi
}\frac{\eta}{(\varepsilon-E_{+})^{2}+\eta^{2}}\right. \nonumber\\ 
&&\left. +\left|  u_{1,-}\right|^{2}\frac{1}{\pi}\frac{\eta}{(\varepsilon-E_{-})^{2}+\eta^{2}}\right\},\end{eqnarray}
where $\delta(\varepsilon-E_{\pm})$ are the Dirac delta functions at $E_{\pm
}=\frac{1}{2}[(E_{1}+E_{2})\pm\hbar\omega]$, with $\hbar\omega=\sqrt
{(E_{1}^{{}}-E_{2})^{2}+4\left|  V_{1,2}\right|  ^{2}.}$ The eigenstates are
$\left|  +\right\rangle =u_{1,+}\left|  1\right\rangle +u_{2,+}\left|
2\right\rangle $ and $\left|  -\right\rangle =u_{1,-}\left|  1\right\rangle
+u_{2,-}\left|  2\right\rangle$ with the coefficients
\begin{equation}
u_{1,\pm}=\left\{  \frac{V_{1,2}}{2\left|  V_{1,2}\right|  ^{2}}\left[
1\pm\frac{E_{1}-E_{2}}{\hbar\omega}\right]  \right\}  ^{1/2},
\label{eq_1st2x2}%
\end{equation}%
\begin{equation}
u_{2,\pm}=\pm\left\{  \frac{V_{1,2}}{2\left|  V_{1,2}\right|  ^{2}}\left[
1\mp\frac{E_{1}-E_{2}}{\hbar\omega}\right]  \right\}  ^{1/2}.
\label{eq_2st2x2}%
\end{equation}
An immediate advantage of the Green's function formalism is that
although it is obtained through a finite number of algebraic
operations (matrix inversion) it contains information on both the
eigenenergies and eigenfunctions which involve the transcendental
operation of finding the roots of a polynomial. In particular, any 
eigenvector component is obtained from
the generalization of Eq. (\ref{eq_LDOS2}) which results:
\begin{equation}
N_{i}(\varepsilon)    =  -\frac{1}{\pi}\lim_{\eta \rightarrow 0}
{\rm Im}G_{i,i}^{R}(\varepsilon+{\rm i}\eta)\label{eq_LDOS}\\
\end{equation}

\begin{figure}[tbp]
\narrowtext
\centering \leavevmode
\center{\epsfig{file=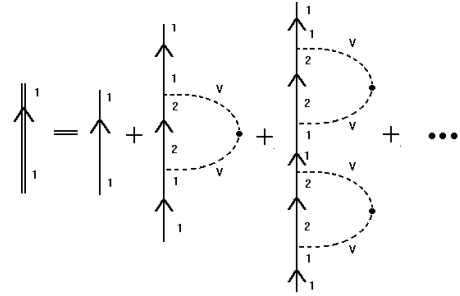,width=6.3cm}} 
\vspace{0.5cm}
\caption{Feynman diagrams in tight-binding representation. Double and 
simple lines represent the exact and unperturbed Green's function
respectively. Hopping interactions are represented by the dashed lines
that modifies the indices of a Green's function. Two conjugate
interactions are associated by the dot into a single rung.}
\label{fig_fey-Gseries}
\end{figure}

The connection with the infinite order perturbation theory is immediate just
expressing the denominator in Eq. (\ref{eq_Green2x2}) through its series
expansion
\begin{eqnarray}
G_{1,1}^{R}(\varepsilon)  &  = &\frac{1}{\varepsilon+{\rm i}\eta
-E_{1}-V_{1,2}\displaystyle\frac{1}{\varepsilon+{\rm i}\eta-E_{2}}V_{1,2}
},\label{eq_inverse-exp}\\
&  = &\frac{1}{\left[  G_{1,1}^{(o)R}(\varepsilon)\right]^{-1}-V_{1,2}
G_{2,2}^{(o)R}(\varepsilon)V_{2,1}}\\
&  = & G_{1,1}^{(o)R}(\varepsilon)[1-\sum_{n=0}^{\infty}\left(V_{1,2}
G_{2,2}^{(o)R}(\varepsilon)V_{2,1}\right)  ^{n}]\label{eq_G11-1/(1-x)}\\
&  = & G_{1,1}^{(o)R}(\varepsilon)\nonumber\\
&&+G_{1,1}^{(o)R}(\varepsilon)V_{1,2}
G_{2,2}^{(o)R}(\varepsilon)V_{2,1}G_{1,1}^{(o)R}(\varepsilon)\nonumber\\
&&+\cdots,
\label{eq_pert-series}\\
&  = & G_{1,1}^{(o)R}(\varepsilon)+G_{1,1}^{(o)R}(\varepsilon)\Sigma
_{1,1}(\varepsilon)G_{1,1}^{R}(\varepsilon) \label{eq_Dyson}
\end{eqnarray}
The last line corresponds to the Dyson equation while the series
expansion is represented diagrammatically in
Fig.\ref{fig_fey-Gseries}.

In systems with a finite number, $N,$ of states $G_{i,j}(\varepsilon)$
is a well behaved meromorphic function except at the $N$ poles. In
this condition, it is satisfied that
\begin{equation}
{\rm finite\,\,system}\left\{
\begin{array}
[c]{c}%
\Gamma(\varepsilon)=-{\rm Im}[\Sigma(\varepsilon)]\equiv0,\\
\Sigma(\varepsilon)\equiv\Delta(\varepsilon)={\rm Re}[\Sigma
(\varepsilon)].
\end{array}
\right.  \label{eq_reality}%
\end{equation}
Therefore, by considering real energies we can drop the supra-index that
distinguish the retarded and advanced Greens function.

Let us come back to the three site model. We may sum up the infinite
perturbation series to obtain
\wide{m}{
\begin{equation}
G_{1,1}(\varepsilon)=\displaystyle\frac{ 1}{\varepsilon-E_{1}-V_{1,2}\displaystyle\frac{1}%
{\varepsilon-E_{2}+V_{2,3}\displaystyle\frac{ 1}{ \varepsilon-E_{3}}V_{3,2}}V_{2,1}%
-V_{1,3}\displaystyle\frac{ 1}{ \varepsilon-E_{3}}V_{3,1}}, \label{eq_G11_3x3}%
\end{equation}
}
or equivalently use Eq. (\ref{eq_3x3-->2x2paramet}) to regrouping the terms
into:
\begin{equation}
G_{1,1}(\varepsilon)=\frac{1}{\varepsilon-\widetilde{E}_{1}-\widetilde{V}_{1,3}%
\displaystyle\frac{1}{\displaystyle \varepsilon-\widetilde{E}_{3}}\widetilde{V}_{3,1}}, \label{eq_G11_3x3short}%
\end{equation}
which gives the obvious connection with the real space decimation procedure.

Many other useful recursion formulas can be obtained by similar procedures.
For example, in a system with states $\{0,1,2,3,\cdots N,N+1\}$, the effective
hopping is related to the non-diagonal Green's function in the isolated bridge
with states $\{1,2,3,\cdots N\},$
\begin{eqnarray}
\widetilde{V}_{0,N+1}  &  =  &\widetilde{V}_{0,N}\left[\varepsilon-\widetilde{E}
_{N}\right]  ^{-1}V_{N,N+1},\label{eq_Veff(G1N)}\\
 =  V_{0,1}& &\left\{{G}_{1,N-1}V_{N-1,N}\left[\varepsilon
-E_{N}-\Delta_{N}^{-}\right]^{-1}\right\}  V_{N,N+1},\nonumber\\
& = & V_{0,1}\left\{{G}_{1,N}\right\}  V_{N,N+1}.
\end{eqnarray}
This equation is complemented with the ones for the self-energies from
sites at right
\begin{equation}
\Delta_{0(N)}^{+}(\varepsilon)=V_{0,1}{G}_{1,1}V_{1,0}
\label{eq_Delta11(G11)},%
\end{equation}%
and at left
\begin{equation}
\Delta_{N+1(N)}^{-}(\varepsilon)=V_{N,N+1}{G}_{N,N}V_{N,N+1}
\label{eq_DeltaNN(Gnn)}.%
\end{equation}
One can then calculate by iteration of the non-diagonal Greens
Function in progressively bigger systems $\{0,\cdots N+2\},$
$\{0,\cdots N+3\},$ and so on, with an equation that is obtained
equating the terms between brackets in Eq.(\ref{eq_Veff(G1N)}):
\begin{equation}
G_{1,N+1}^{(N+1)}=G_{1,N}^{(N)}V_{N,N+1}G_{N+1,N+1}^{(N+1)},
\label{e_G-nondiag-rec}%
\end{equation}
which is used in conjunction with,
\begin{equation}
G_{N+1,N+1}^{(N+1)}=\left[  \varepsilon I-E_{N+1}-V_{N+1,N}G_{N,N}%
^{(N)}V_{N+1,N}\right]  ^{-1}, \label{eq_G-diag-rec}
\end{equation}
the continued fraction expansion of the diagonal Greens function.

\subsection{Going beyond one-dimensional systems.}

\begin{figure}[tbp]
\narrowtext
\centering \leavevmode
\center{\epsfig{file=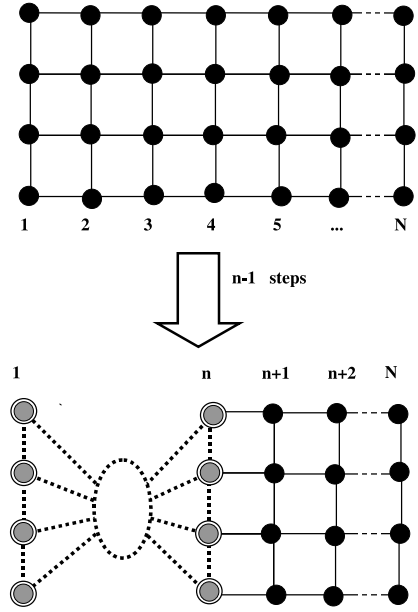,width=5.5cm}} 
\vspace{0.5cm}
\caption{Scheme of the decimation of a finite tight-binding strip. As the
elimination of intermediate layers proceeds the effective site
energies and interaction appears.}
\label{fig_dec-red}
\end{figure}

The decimation scheme can be generalized to any dimension as long as we proceed
in a  ``layer by layer'' elimination. Every site in the above procedure now
becomes an $n\times n$-matrix where $n$ is the size of the layer. In Fig.
\ref{fig_dec-red} we consider a square lattice of orbitals forming a
strip. Remembering that ${\rm Im}[{\mathbf{\Sigma}}]\equiv0$ we write the
general equation for the real self-energies. We proceed to eliminate
one by one the layers from 1 to the ($N-1$). Adopting the matrix
notation ${\mathbf {\Delta}}_{1(n)}^{+}(\varepsilon)$ is the self-energy correction to the
$1^{\rm st}$ layer when all layers (to the right) including the
$n^{\rm th}$ have been eliminated. ${\mathbf{\Delta}}_{n+1}^{-}(\varepsilon)$
is the self-energy correction to layer ($n+1$) when layers to the left
have been eliminated.  ${\mathbf{\widetilde{V}}}_{1,n+1}$ is the effective
interaction between layers.  These are again evaluated with the
iterative procedure
\begin{equation}
{\mathbf{\Delta}}_{1(n)}^{+}(\varepsilon)={\mathbf{\Delta}}_{1(n-1)}
^{+}+{\mathbf \widetilde{V}}_{1,n}\frac{1}{\varepsilon {\mathbf I}-{\mathbf E}
_{n}-{\mathbf{\Delta}}_{n}^{-}}{\mathbf\widetilde{V}}_{n,1}, 
\label{eq_add_S}
\end{equation}%

\begin{equation}
{\mathbf{\Delta}}_{n+1}^{-}(\varepsilon)={\mathbf V}_{n+1,n}\frac{1}
{\varepsilon{\mathbf I}-{\mathbf E}_{n}-{\mathbf{\Delta}}_{n}^{-}}{\mathbf V}_{n,n+1}, 
\label{eq_MCF}
\end{equation}
\begin{equation}
{\mathbf \widetilde{V}}_{1,n+1}={\mathbf \widetilde{V}}_{1,n}\frac{1}{\varepsilon
{\mathbf I}-{\mathbf E}_{n}-{\mathbf{\Delta}}_{n}^{-}}{\mathbf V}_{n,n+1}
+{\mathbf V}_{1,n+1}, 
\label{eq_eff-V}
\end{equation}
with the initial values:
\begin{equation}
\begin{array}
[c]{c}%
{\mathbf{\Delta}}_{1}^{+(1)}(\varepsilon)={0,}\\
{\mathbf{\Delta}}_{2}^{-}(\varepsilon)={0,}\\
{\mathbf{\widetilde{V}}}_{1,2}={\mathbf{V}}_{1,2}.
\end{array}
\label{Eq_intitial-matrices}
\end{equation}
The expression one has to evaluate are expressed as Matrix Continued
Fractions, which are numerically very stable \cite{z--Pasta-MCF+G}. Notice
that the elimination of intermediate layers produces effective interaction
among the sites in the first layer, i.e. it modifies the intra-layer
interactions as prescribed by the non-diagonal elements of ${\mathbf{\Delta}}$.
Inter-layer interactions are always contained in $\mathbf{V}$. Again, the last
term in Eq. (\ref{eq_eff-V}) allows the consideration of interactions going beyond nearest neighbor layers.

\subsection{Decimation in Molecules}

\begin{figure}[tbp]
\narrowtext
\centering \leavevmode
\center{\epsfig{file=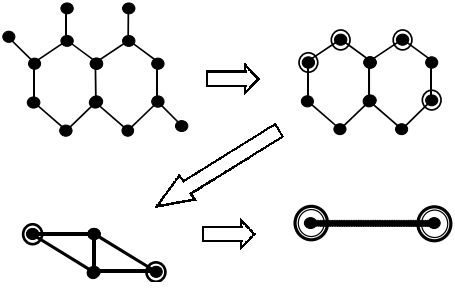,width=6.0cm}} 
\vspace{0.5cm}
\caption{Sequence of the decimation of a molecule into an effective two 
site problem.}
\label{fig_deci-molecule}
\end{figure}

It is clear that the described procedure is very well suited for
application to molecular systems\cite{z--Levstein-decim,z--Ratner}. 
Just to fix ideas consider the $p_{z}$ orbitals in the carbon backbone 
of an organic molecule represented in Fig.\ref{fig_deci-molecule}. Assume that we
are interested in the study of how charge can be transferred from site
$L$ to site $R$, a typical problem in photosynthesis and molecular
electronics. Circles represent the $\pi$ orbitals with given local
energies, lines are hopping interactions which produce the electron
delocalization. One might start decimating the dangling ends. Sites
energies in the back-bone are then renormalized. Next, we identify the
branching nodes and eliminate the bridging sites, this gives new site
energies and an effective hopping. Finally one eliminates all the
remaining bridging structure obtaining an effective two orbital molecule. 
 Since the procedure is exact one obtains the exact spectrum independently 
of the election of  these orbitals. 
However, in transport one wants to keep the atoms that
have matrix elements that enable the transfer of charge with the
``external world''. This will become clearer later.

The spectral and transport properties of model molecules are further
discussed in Ref.(\cite{z--Levstein-decim}). Let us mention that the
procedure allows to visualize a situation of maximum coupling when
$\widetilde{E}_{L}=\widetilde{E}_{R}$ which is a resonant situation. This
paper also discusses for the first time a situation of maximum
decoupling which occurs when $\widetilde{V}_{L,R}=0$ and was named
``antiresonance'' or minimal effective decoupling of the centers. This phenomenon
is caused by the interference between different pathways when
energies of the pathway molecules lie between energies of the
other. This generalizes, to the transport case, a concept introduced by
Fano\cite{z--Fano-old} in ionization spectroscopy.

\subsection{Advantages of the decimation procedure}

One might think that instead of decimating a particular structure one
might invert $(\varepsilon\mathbf{I}-\mathbf{H})$ directly. However,
there are various advantages in favor of the decimation procedure
explained above:

a) The scheme adapts naturally to increase the size of the system
without having to recalculate the matrices again.

b) The sparse nature of the Hamiltonian is naturally included in the recursion
formula saving storage memory and iteration steps.

c) The decimation scheme has implicit a deep physical insight of the
system and reveals the self-similarities of the systems whenever those
properties are present.

d) The resulting highly singular resolvent of a big system through
matrix inversion is in general numerically unstable. In contrast, the
decimation, is a numerically stable procedure. The natural
instabilities of the system such as exponentially growing
eigensolutions are explicitly used in favor of the convergence of the
method. For further discussion on this issue see
Ref.\cite{z--Pasta-MCF+G}. On this basis one establishes that Eq.
(\ref{eq_MCF}) which defines a Matrix Continued Fraction procedure is
more accurate and stable than Eq.(\ref{eq_G-diag-rec}) which uses a
different iteration procedure to calculate an equivalent magnitude.

\subsection{The Thermodynamic limit: Dyson Equation}

We have seen that any finite Hamiltonian matrix of finite size can be solved
through the decimation method obtaining a set of discrete eigenstates. One
might wonder what happens when the number of orbital is actually infinite. How
and when does the continuum spectrum appear?. Let's come back to the
one-dimensional model. The decimation procedure allows to deal with a simple
but already non-trivial case: that of an ordered semi-infinite chain. In this
situation: $V_{n,n+1}=V$ and $E_{n}\equiv E_{0}$ for every $n.$ The Dyson
equation is:
\begin{equation}
\Sigma_{n}=V\frac{1}{\varepsilon-E_{0}-\Sigma_{n+1}}V, \label{eq_Sigma+Dyson}%
\end{equation}
and must include the simple fact that every site sees to the right an infinite
chain: $\Sigma_{n}\equiv\Sigma_{{}}$ for every $n$. That is a \ new way to
present the Bloch theorem.%

\begin{equation}
\Sigma^{{}}=V\frac{1}{\varepsilon-E_{0}-\Sigma^{{}}}V=\Delta\mp{\rm i}\Gamma. 
\label{eq_Bloch-->Dyson}
\end{equation}
The striking fact is that even when we are working\ with real
quantities the solution of this equation may lay in the complex
plane\cite{z--Weisz-Pastawski}. There are two possibilities for the
imaginary part. We call retarded self-energy to that which would cause
a decay in the time evolution of the wave function amplitude. The
solution of the second order equation gives:
\wide{m}{
\begin{equation}
\Delta=\left\{
\begin{array}
[c]{cc}%
\frac{\varepsilon-E_{0}}{2}-\sqrt{\left(  \frac{\varepsilon-E_{0}}%
{2}\right)  ^{2}-V^{2}} & {\rm if\,\,}\varepsilon-E_{0}>2\left|  V\right|
,\\
\frac{\varepsilon-E_{0}}{2} & {\rm if\,\,}\left|  \varepsilon
-E_{0}\right|  \leq2\left|  V\right|  ,\\
\frac{\varepsilon-E_{0}}{2}+\sqrt{\left(  \frac{\varepsilon-E_{0}}%
{2}\right)  ^{2}-V^{2}} & {\rm if\,\,}\varepsilon-E_{0}<-2\left|  V\right|
,
\end{array}
\right.  \label{eq_Delta-1d}%
\end{equation}
}
and
\begin{equation}
\Gamma=\left\{
\begin{array}
[c]{cc}%
0 & {\rm if\,\,}\left|  \varepsilon-E_{0}\right|  >2\left|  V\right|  ,\\
\sqrt{\left( V^{2}- \displaystyle\frac{\varepsilon-E_{0}}{2}\right)  ^{2}} &
{\rm if\,\,}\left|  \varepsilon-E_{0}\right|  \leq2\left|  V\right|  .
\end{array}
\right.  \label{eq_Gamma_1d}%
\end{equation}
We see that in the region were the Block solutions occur, i.e. when
$\varepsilon=E_{k}\equiv E_{0}-2V\cos(ka),$ is the region where the
spectrum is absolutely continuous, one has $\Gamma(\varepsilon)\neq0.$
\ In the region of the real axis, the Green's functions
indetermination represents a branch cut, and it becomes non-analytic
at the spectral support. However, the infinite dimensionality of the
Hilbert space is a necessary but not sufficient condition for the
continuum spectrum. For example, for 1-d systems with disorder in the
site energies described by the Anderson model, the self-energies are
always real even when the system is infinite i.e. there is genuine
``phase transition'' in the nature of the states when one goes from
finite systems to infinite ones. In an infinite system, disorder can
produce a localized-extended transition \cite{z--MacKinnon-review}
first described by P. W. Anderson and commonly referred as
metal-insulator transition (MIT). The thermodynamic limit that makes
possible the study of such transition is to study the observable
$\mathcal{O}$ (decay rate, density of states, etc.) in the limit
\begin{equation}
\lim_{\eta\rightarrow0}\lim_{N\rightarrow\infty}\mathcal{O}.
\label{eq_therm-limit}%
\end{equation}
The order of this limit was implicit when we searched for complex solutions
for the Dyson equation by imposing the ``Bloch theorem''.

We must remember that the MIT is not exclusive to the disordered
systems. It is also of frequent occurrence in ordered systems under the
action of incommensurate potentials \cite{z--Weisz-Pastawski}. For
example a 1-d systems where the site energy is of the form
$E_{n}=W\cos\left[ Qna\right] $ with $2\pi/Q$ incommensurate with the
lattice length $a$ provided that $W>W_{c}=2V$.

\subsection{Density of States in unbounded Systems: Crystals}

In the case where the chain extends to both sides of site 0 one gets
two contributions to the self energy correction. All diagonal terms of
the Green's function are identical
\begin{eqnarray}
G_{0,0}^{R}& = &\frac{1}{\varepsilon-E_{0}-2\Sigma^{R}},\nonumber\\
& = &\frac{1}{\varepsilon
-E_{0}-2\frac{\left(  \varepsilon-E_{0}\right)  -{\rm i}\sqrt
{4V^{2}-\left(  \varepsilon-E_{0}\right)  ^{2}}}{2}}, \label{eq_Goo-1d}%
\end{eqnarray}
from which the density of states per site (DoS) can be evaluated,
\begin{equation}
N_{\left[  d=1\right]  }\left(  \varepsilon\right)  =\frac{1}{2\pi V}\frac
{1}{\sqrt{1-\left(  \displaystyle\frac{\varepsilon-E_{0}}{2V}\right)  ^{2}}}.
\label{eq_DOS_1d}%
\end{equation} which has the characteristic van Hove singularities of one dimensional systems
at the band edges.

The fact that the imaginary part in the self-energy is associated to the
``irreversible decay'' of the state can be formalized by studying the Fourier
transform of the Greens function to obtain the return probability
\begin{eqnarray}
P_{0,0}(t)  &  = &\left|\left\langle 0\right| \exp\left[-{\rm i}
{\mathcal H}t/\hbar\right ]  \left|  0\right\rangle \right| ^{2} \nonumber\\
& = &\left|  \int G_{0,0}^{R}(\varepsilon)\exp\left[  -{\rm i}{\mathcal\varepsilon}
t/\hbar\right]  \mathrm{d}\varepsilon\right|  ^{2}\label{eq_return-prob}\\
&  = &\left|  J_{0}(tV/\hbar)\right|  ^{2}\sim\frac{1}{\pi}(tV/\hbar)^{-1}.
\label{eq_1-d_quantum_diff}
\end{eqnarray}
where $J_{0}$ is the Bessel function. For long times, it represents an
irreversible ``super-diffusion'', and hence differs substantially from a
classical random walk process in an infinite chain.

\begin{figure}[tbp]
\narrowtext
\centering \leavevmode
\center{\epsfig{file=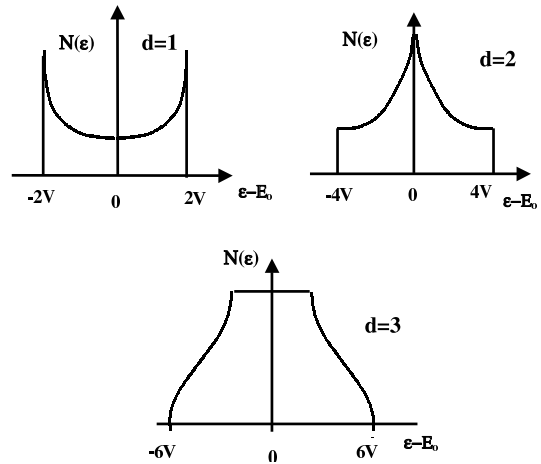,width=7.0cm}} 
\vspace{0.5cm}
\caption{Schematic Density of States in 1-d, 2-d and 3-d systems.}
\label{fig_DoS_1d-2d-3d}
\end{figure}

Densities of states in higher dimensions for hyper-cubic lattices can be
obtained from those in one dimension by ussing the fact that variables are separable.
This allows us to split the energy contributions of each dimension
$E({\bf k})=E_{0}+E(k_{x})+E(k_{y})+E(k_{z}),$ leading to a convolution
expression for the DoS in dimension d=2 and d=3:
\begin{equation}
N_{\left[  2\right]  }\left(  \varepsilon\right)  =\int N_{\left[  1\right]
}\left(  \varepsilon-\varepsilon^{\prime}\right)  N_{\left[  1\right]
}\left(  \varepsilon^{\prime}\right)  {\rm d}\varepsilon^{\prime}
\label{eq_conv-2d}
\end{equation}
and in general
\begin{equation}
N_{\left[  d+1\right]  }\left(  \varepsilon\right)  =\int N_{\left[  d\right]
}\left(  \varepsilon-\varepsilon^{\prime}\right)  N_{\left[  1\right]
}\left(  \varepsilon^{\prime}\right)  {\rm d}\varepsilon^{\prime}.
\label{eq_conv-3d}%
\end{equation}
For example, in a square lattice we get
\begin{equation}
N_{\left[2\right]  }\left(  \varepsilon\right)  =\frac{1}{4}K\left[
\sqrt{1-(\frac{\varepsilon}{2V})^{2}}\right]  , \label{eq_DOS-2d}%
\end{equation}
where $K$ is the complete elliptic integral of the first kind. Fig.
\ref{fig_DoS_1d-2d-3d} show the qualitative features of these DoS. See Ref.
\cite{z--Pastawski-Wiecko} for a more extended discussions on these points and
plots of the DoS in higher dimensions.

Now that we know how to evaluate some basic self-energies, we can
calculate the Green's function in a great variety of model systems.

\subsection{Surfaces in a semi-infinite chain.}

Let us consider the semi-infinite chain $\{s,1,2,3,\cdots\}$. The Green's
function at the surface site is
\begin{equation}
G_{1,1}^{R}=\frac{1}{\varepsilon-E_{0}-\Sigma^{R}(\varepsilon)},
\label{eq_G11}%
\end{equation}
where we have used the subindex $1$ for the Green's function to stress on the
fact that we count orbitals starting at the surface. One gets the density of
states for the surface of the chain
\begin{equation}
N_{{surf.}}\left(  \varepsilon\right)  =\frac{1}{\pi V}\sqrt{1-\left(
\frac{\varepsilon-E_{0}}{2V}\right)  ^{2}.} \label{eq_DOSsurf}%
\end{equation}
Incidentally let us note that an identical DoS is obtained when the
Hamiltonian is a Random Matrix\cite{z--RMT-Mattis}. In this case one
can use the Lanczos method to tridiagonalize the matrix in the
infinite dimension limit and see that it corresponds to the ordered
chain we just discussed .

The return probability to the surface site, as defined in
Eq.(\ref{eq_return-prob}) gives
\begin{equation}
P_{1,1}(t)=\left|  \frac{\hbar}{Vt}J_{1}(2Vt/\hbar)\right|  ^{2}\simeq
\frac{1}{\pi}\left[  \frac{\hbar}{Vt}\right]  ^{3}, 
\label{eq_surface-retur}
\end{equation}
indicating a decay of the surface state.  This exact result, which
differs from the usual exponential and diffusive laws usually adopted
for the empirical description of decay phenomena, describes an
``irreversible'' escape from the site. This contrasts with finite
systems where recurrences, called mesoscopic
echoes\cite{z--mesoscopic-echoes,z--cook-AFA}, appear at a typical
time estimated as $\hbar/\Delta,$ called the Heisenberg time.

Perhaps, the most important conclusion to be drawn from the previous
results is the correction to energies of finite systems to account for
its contact to infinite systems.

\subsection{Adatoms and Surface states.}

The method given above allows us to extend the decimation procedure for finite
systems described in the previous section to composite systems (finite
+infinite). Let us discuss the simplest but still non-trivial example. Indeed
\ one can think of the 0-th site as an \textit{adatom} in the surface of a
metal. In that case its energy $E_{s}$ and hopping element $V_{s}$ are
different from those of the bulk, $E_{0}$ and $V$ respectively. The
Green's function evaluation leads to
\begin{equation}
G_{s,s}^{R}=\frac{1}{\varepsilon-E_{s}-\left|  \frac{V_{s}}{V}\right|
^{2}\frac{\left(  \varepsilon-E_{0}\right)  -{\rm i}\sqrt{4V^{2}-\left(
\varepsilon-E_{0}\right)  ^{2}}}{2}}. \label{eq_Gss}%
\end{equation}
We leave to the reader the evaluation of the local DoS both
numerically and analytically, as well as the calculation of the
condition for the appearance of localized states and resonances. For $V_{s}=V,$  a localized state may appear. It shows up as an isolated pole in $G_{s,s}^{R}(\varepsilon)$  in the real axis at 
\begin{equation}
{\widetilde E}_{s}=\left( \left[E_s-{E}_{0}\right]^2
+V^2\right)/\left[E_s-E_0\right],
\label{eq_local-adatom}
\end{equation}
provided that $\left| -
E_{0}\right|>0$.

One can also check that for $\left|  E_{s}-E_{0}\right|  <V$ and $\left|
\frac{V_{s}}{V}\right|  \ll 1,$ a resonant state will appear a pole in the complex plane with a real part overlaping with continuous band. Its energy is:
\begin{equation}
\widetilde{E}_{s}\simeq E_{s}+\left|  \frac{V_{s}}{V}\right|  ^{2}\Delta(E_{s}).
\label{eq_energy-shift}%
\end{equation}
The return probability is well approximated by an exponential law as
prescribed by the Fermi Golden Rule:
\begin{eqnarray}
P_{s,s}(t)  &  \sim & \exp[-t/\tau_{s}],\label{eq_return-adatom}\\
{\rm with}\,\,\,\,1/\tau_{s}  &  = &\frac{2\pi}{\hbar}\left|  V_{s}\right|
^{2}N_{{\rm surf}.}(E_{s})\simeq 2 \Gamma_s(E_s)/\hbar. 
\label{eq_decay-adatom}
\end{eqnarray}
Again, one sees that the time dependencies that appear in Quantum Mechanics can
be very different from those present in a classical random walk.

\subsection{Localized state in a branched circuit.}

There is a highly non-trivial result\cite{z--DAmato AB} that can be easily
obtained with the help of the above formalism. This is the existence of
localized states in branching regions.

In general, for a site $x$ to which $z$ semi-infinite chains are
connected, the Green's function at the crossing site between branches
can be written as
\begin{equation}
G_{x,x}^{R}=\frac{1}{\varepsilon-E_{0}-z \ \displaystyle\frac{\left(  \varepsilon
-E_{0}\right)  -{\rm i}\sqrt{4V^{2}-\left(  \varepsilon-E_{0}\right)  ^{2}%
}}{2}}. 
\label{eq_Gxx-BRANCH}
\end{equation}
Which besides of the expected branch cut in $\left|  \varepsilon-E_{0}\right|
<2V$\thinspace, it presents isolated poles at
\begin{equation}
E_{{\rm branch}}=E_{0}\pm 2V\frac{z^{2}}{4(z-1)} \label{eq_E-local-BRANCH}%
\end{equation}
These solutions, are not only \ present in a branching polymer, but in a
conveniently grown heterostructure \cite{z--T-shaped/Esaki}. The existence of
a topologically confined states in the case $z=3$ has become the basis for
quantum transistors \cite{z--T-transistor}\ and lasers\cite{z--Pinczuk}.

\subsection{Representation of the environment through self-energies}

First we recall that one can always eliminate microscopic degrees of freedom
\cite{z--Löwdin,z--Levstein-decim} generating an effective Hamiltonian that
accounts for them exactly. This produces effective interactions and energy
renormalization which depend themselves on the observed energy. Furthermore,
one can include the effects of a whole lead in a Hamiltonian description
through an \textit{imaginary} correction to the eigenenergies. In fact, an
electron originally localized in the region called ``the sample'' should
eventually escape or decay toward the lead. Hence, there is a escape velocity
associated with the energy uncertainty of a local state $i$:%
\begin{equation}
v_{i}=\frac{2a}{\hbar}\Gamma_{i}=\frac{a}{\tau_{i}}. 
\label{eq_v-Gamma}
\end{equation}
Other interactions that one usually considers produce decay rates
reasonably well described by the Fermi Golden Rule: electrons in an
excited atom decaying into the continuum, or propagating electrons
decay into different momentum states by collision with impurities
producing phonons or photons. In some of these cases we have to add
some degrees of freedom to the sum (the phonon or photon
coordinates). A process $\alpha$ may produce a complex contribution
$^{\alpha }\Sigma_{0}^{R}$ which adds into the total self-energy
$\Sigma_{0}^{R}$. In particular this is the interpretation one 
must give to the small imaginary part $\eta$ we introduced in the 
definition of the Green's function of Eq. (\ref{eq_G-matrix}). 
The effective Hamiltonian becomes: 
\begin{equation}
\hat{\mathcal{H}}^{(o)}{}_{\overrightarrow{{\rm interactions}}}%
\hat{{\mathcal H}}=\hat{{\mathcal H}}^{(o)}+\hat{\Sigma}^{R},
\label{eq_perturbed-Ho}%
\end{equation}
with
\begin{equation}
\hat{\Sigma}_{{}}^{R(A)}
=\sum_{i}({\Delta}_i \mp{\rm i}{\Gamma}_i )(\left|  i\right\rangle
\left\langle i\right| . \label{eq_def-Sigma}%
\end{equation}
That is, we can not find eigenstates because the complete Hamiltonian is not
separable into a product of states of the unperturbed system and those of the
environment. 

\begin{figure}[tbp]
\narrowtext
\centering \leavevmode
\center{\epsfig{file=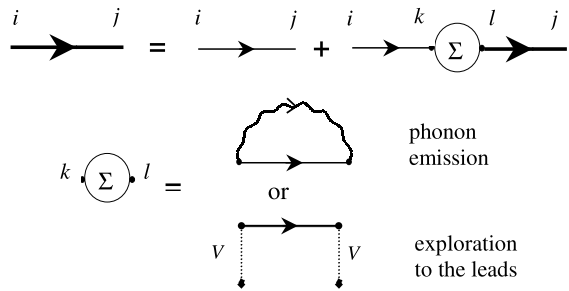,width=7.5cm}} 
\vspace{0.5cm}
\caption{Diagrams shown before can be rearranged into the self-energy correction
$\Sigma$. This equation is valid for independently of the approximation used
to calculate $\Sigma$. In the lower panel self-energies obtained through
escape to the leads and electron-phonon interactions are shown. The escape
self-energies contain the hopping (dot) and a propagator in the lead. An
electron-phonon self-energy is evaluated in terms of the sample's electron
(straight-line) and phonon (wavy-curve) Green's functions.}
\label{fig_Feynman-G-Sigma}
\end{figure}

In terms of the Green's function the resulting equation is, in
the matrix representation:
\begin{equation}
{\bf G}^{R}={\bf G}^{(o)R}+{\bf G}^{(o)R}{\bf\Sigma}_{{}}
^{R}{\bf G}^{R}. 
\label{eq_Dyson-Matrix}
\end{equation}
The diagrammatic expansion for the perturbed Green's function is shown in Fig.
\ref{fig_Feynman-G-Sigma}.

Although the Green's function formalism seems to introduce some extra
notation, this has various conceptual advantages. For example, it is
straightforward to use Eq.(\ref{eq_Dyson-Matrix}) to prove a version of the \textit{optical
theorem} which relates the imaginary part of the forward scattering amplitude with the total cross section of the pertubation. One starts by writting Eq. (\ref{eq_Dyson-Matrix}) in the form:
\begin{equation}
{\bf\Sigma}^{R}
   = 
({\bf G}^{(o)R})^{-1}-({\bf G}^{R})^{-1},
\label{eq_Sigma-Dyson-matrix}
\end{equation}
and an equivalent one for ${\bf\Sigma}^{A}$. The difference between them is:
\begin{eqnarray}
{\bf\Sigma}^{R}-{\bf\Sigma}^{A}
& = &({\bf G}^{A})^{-1}-({\bf G}^{R})^{-1},\nonumber\\
& = &({\bf G}^{R})^{-1}\left[{\bf G}^{R}-{\bf G}^{A}\right]({\bf G}^{A})^{-1},
\label{eq_Imag-Sigma-matrix}
\end{eqnarray}
from which we obtain:
\begin{equation}
\left[  {\mathbf G}^{R}-{\mathbf G}^{A}\right]  ={\mathbf G}^{R}\,\,\left[
{\mathbf{\Sigma}}^{R}-{\mathbf{\Sigma}}^{A}\right]  \,\,\,\mathbf{G}^{A},
\label{eq_opticaltheorem}
\end{equation}
which has deep physical significance. It is an integral equation relating the
local densities of states given Eq.(\ref{eq_LDOS}) and the decay rates
provided by Eq. (\ref{eq_v-Gamma}). Perhaps, the most important advantage of
Green's functions, is that they can be used also in the Quantum Field Theory
\cite{z--Keldysh,z--Danielewicz} to deal with the many-body case and are
a relevant tool in the clarification of the problem of tunneling time.

\section{The Fisher and Lee formula}

The original demonstration \cite{z--Fisher-Lee} of the connection between
transmittance and Green's function is based on properties of the scattering
matrix in the continuum model. It is also presented in a simplified form in
the book by Datta\cite{z--Datta}. Here we are going to present the main lines
of a demonstration based on properties of a molecular orbital model in its
decimated tight binding version.

The isolated sample can be decimated into two sites getting an exact
non-linear effective Hamiltonian:%
\wide{t}{
\begin{equation}
{\hat{\mathcal H}}_{0}={E}_{1}\left|  1\right\rangle \left\langle 1\right|
+{E}_{N}\left|  N\right\rangle \left\langle N\right|  +{V}
_{1,N}\left|  1\right\rangle \left\langle N\right|  +{V}_{N,1}\left|
N\right\rangle \left\langle 1\right|  , \label{eq_Ho}%
\end{equation}
where the parameters ${E}$ and ${V}$ are functions of the
variable $\varepsilon.$

We attach electrodes or leads at each site. Both leads are perfect
semi-infinite 1-d chains where the dispersion relations are
\begin{equation}
^{{}}E_{k{L}}=E_{{L}}+2V_{{L}}\cos[k_{{L}}a],
\label{eq_disperssion-relation}
\end{equation}
from which the group velocity is
\begin{equation}
v_{{L}}=\frac{1}{\hbar}\frac{\partial E_{k{L}}}{\partial
k_{{L}}}=-\frac{2aV_{{L}}}{\hbar}\sin[k_{{L}}a].
\label{eq_velocity-Gamma_chain}%
\end{equation}
Analogous relations for $E_{k{R}}$ and $v_{{R}}$ hold in the
right channel. When the perturbation
\begin{equation}
\hat{{\mathcal H}}_{0-{\rm leads}}=V_{t{L}}(\left|  1\right\rangle
\left\langle 0\right|  +\left|  0\right\rangle \left\langle 1\right|
)+V_{t{R}}(\left|  N\right\rangle \left\langle N+1\right|  +\left|
N+1\right\rangle \left\langle N\right|  ) \label{eq_Ho-leads}
\end{equation}
is turned on, the new single particle eigenfunctions components

\begin{equation}
u_{k}(an)=\left\{
\begin{array}
[c]{c}
\exp[{\rm i}kna]+r\exp[-{\rm i}kna]\\
A\\
B\\
t\exp[{\rm i}kna]
\end{array}
\right.  \,\,\,\,\,\,\,
\begin{array}
[c]{c}
{\rm if\,}\,\,\,n\leq0,(na\in{L}),\\
{\rm if\,}\,\,\,n=1,\\
{\rm if\,}\,\,\,n=N,\\
{\rm if\,}\,\,\,\,\,n\geq N+1,(na\in{R}),
\end{array}
\label{eq_TB-wave}%
\end{equation}
are determined from the stationary Schr\"{o}dinger equation 
$\left(
\varepsilon {\hat{\mathcal I}}
-{\hat{\mathcal H}} \right)  \left|  \psi\right\rangle =0.$ We
have only four non trivial terms which contain either $A$ or $B.$ One
eliminates $A$, $B$ and $t$ and evaluates
\begin{eqnarray}
T_{{R,L}}  &  = & 1-r^{\ast}r\\
&  = & 4\left|  \frac{V_{t{L}}}{V_{{L}}}\right|^{2}\left(
\left|  V_{{L}}\right|  \sin{\small [k}_{{L}}a{\small ]}\right)
\left|  {V}_{1,N}\right|  ^{2}\left|  \frac{V_{t{R}}
}{V_{{R}}}\right|^{2}\left( \left| V_{{R}}\right|
\sin[k_{{R}}a]\right)  /[dd^{\ast}],\nonumber\\
\end{eqnarray}
where
\begin{eqnarray}
d  &  = &\left[  \varepsilon-{E}_{1}+\left|  \frac
{V_{t{L}}}{V_{{L}}}\right|  ^{2} \{\left|  V_{{L}}\right|
\cos[k_{{L}}a]  -{\rm i}\left|  V_{{L}}\right|
\sin[k_{{L}}a]\}\right] \nonumber\\
&  \times & \left[  \varepsilon-{E}_{N}+\left|  \frac
{V_{t{R}}}{V_{{R}}}\right|  ^{2} \{\left|  V_{{R}}\right|
\cos[k_{{R}}a]  -{\rm i}\left|  V_{{R}}\right|
\sin[k_{{R}}a]\} \right] \nonumber\\
&  - & \left[ {V}_{1,N}{V}_{N,1}\right].\nonumber
\end{eqnarray}
}
In order to compare this with the Green's function we remember that the
effective site energies include the self-energies produced by the leads
\begin{eqnarray}
^{{L}}\Sigma_{1}  &  = & \left|  \frac{V_{t{L}}}{V_{{L}}
}\right|^{2} \left ( \left|  V_{{L}}\right|  \cos[k_{{L}
}a]-{\rm i}\left|  V_{{L}}\right|  \sin[k_{{L}}a]\right )
\label{eq_Sigma_L-TB}\\
&  = & \,^{{L}}\Delta_{1}-{\rm i\,\,}^{{L}}\Gamma_{1},\nonumber
\end{eqnarray}
and
\begin{eqnarray}
^{{R}}\Sigma_{1}  &  = &\left|  \frac{V_{t{R}}}{V_{{R}}
}\right|  ^{2}  \left(  \left|  V_{{R}}\right|  \cos
[k_{{R}}a]-{\rm i}\left|  V_{{R}}\right|  \sin[k_{{R}
}a]\right)  \\
&  = &\,^{{R}}\Delta_{N}-{\rm i\,\,\,}^{{R}}\Gamma
_{N}.\nonumber
\end{eqnarray}
If the lead variables are decimated they produce an effective potential of the
form
\[
{\hat{ \mathcal{ H}}}_{\mathrm eff.}={\hat{\mathcal {H}}}_{0}+^{{L}}\Sigma_{1}\left|
1\right\rangle \left\langle 1\right|  +^{{R}}\Sigma_{N}\left|
N\right\rangle \left\langle N\right|  .
\]
This maintains the structure of an effective two site Hamiltonian. From this
it is simple to compute the four components of the exact Green's function of
the system
\begin{eqnarray}
G_{1,N}^{R}  &  = &\left\langle 1\right|  \left[  \varepsilon{\hat{\mathcal
I}}-{\hat{\mathcal H}}_{\rm eff.}\right]  ^{-1}\left|  N\right\rangle \\
&  = &\frac{{V}_{1,N}}{\left[  \varepsilon-\left({E}
_{1}+^{{L}}\Sigma_{1}^{R}\right)  \right]  \left[  \varepsilon-\left(
{E}_{N}+^{{R}}\Sigma_{N}^{R}\right)  \right]  -{V}
_{1,N}{V}_{N,1}}\nonumber
\end{eqnarray}
obtaining the Fisher-Lee formula

\begin{equation}
T_{{R,L}}(\varepsilon)=\left(  \frac{\hbar}{a}\right)  ^{2}%
v_{{R}}\,v_{{L}}G_{{R,L}}^{R}(\varepsilon
)\,G_{{L,R}}^{A}(\varepsilon). 
\label{eq_Fisher&Lee}
\end{equation}
Originally, Fisher and Lee considered only the escape velocity to the leads
(i.e. ${\alpha=\beta}={\rm lead)}$. D'Amato and
Pastawski\cite{z--D'Amato-Pastawski} were the first to realize that because of
Eq. (\ref{eq_velocity-Gamma_chain}) one could write transmittances in terms of
the imaginary part of the self energies. Immediately, our point is that
\textit{any other process} which contributes to decay \textit{giving an
imaginary} contribution to the \textit{self-energy} would\ \ be described by
Eq. ($\ref{eq_Fisher&Lee}$). In particular, this will be true for a
\ ``decoherence'' velocity as would be the case of the electron-phonon rate
described in the previous section. In a modern notation:
\begin{equation}
T_{{\alpha R,\beta L}}(\varepsilon)=\left[  2\,^{{\alpha}}%
\Gamma_{{R}}(\varepsilon)\right]  \,G_{{\alpha R,\beta L}}%
^{R}(\varepsilon)\,\left[  2\,^{{\beta}}\Gamma_{{L}}%
(\varepsilon)\right]  G_{{\beta L,\alpha R}}^{A}(\varepsilon).
\label{eq_Fisher-Lee+DP}%
\end{equation}
The left supra-index in $\Gamma$\ indicates the process or channel associated
to the electron decay from the spatial region identified with the right
subindex. Both indices appear as sub-indices in the Green's function. Notice
that $^{{\beta}}\Gamma_{{L}}(\varepsilon),$ which corresponds to
the interaction with the ``source'' of particles, was arranged between two
Green's functions. $^{{\alpha}}\Gamma_{{R}}(\varepsilon),$
representing the properties of the ``sink'', was placed at left. This order is
important in the matrix representation.

\subsection{An example: branched circuit.}

In order to apply the Fisher-Lee formula to a simple but highly non-trivial
example, let us consider again the circuit with $z$ equivalent branches. One
might think that a plane wave incoming at the node from one of the branches
will have a probability $1/(z-1)$ to be transmitted to each of the others,
i.e. that it will behave as a perfect $(z-1)$-splitter. The true answer comes
from the quantum evaluation giving
\begin{equation}
T=\frac{4}{z^{2}}\frac{1}{1+(\frac{2-z}{z})^{2}\displaystyle\frac{(\varepsilon-E_{0}%
)^{2}}{(4V)^{2}-(\varepsilon-E_{0})^{2}}}. \label{eq_T-BRANCH}%
\end{equation}
$T$ has a maximum of $4/z^{2},$ lower than the classical value. Additionally,
the transmittance goes to zero as the energy approaches the band edge. This is
due to the presence of the localized state found in Eq.
(\ref{eq_E-local-BRANCH}) which precludes the occupation of the branching
region by the propagating states.

\subsection{An extension: the multichannel case.}

In fact, one can see that it is not difficult to generalize Eq.
(\ref{eq_G-landauer}) to any number of incoming and outgoing channels
connected to their respective reservoirs. When the two wires
${L}$ and ${R}$ have finite cross section, each wire
contains an integer number, $^{{L}}M_{(\varepsilon_{F})}$ and
$^{{R}}M_{(\varepsilon_{F})},$ of propagating (one-dimensional)
channels (i.e. transversal modes) allowed by the Fermi energy. Each
electronic excitation leaving the wire ${L}$
\ in the propagating channel $i$ with velocity $v_{{L}i}\neq0$, enters
the wire ${R}$ in the propagating channel $j$ with velocity
$v_{{R}j}\neq0$. The conductance in the linear response regime is
\begin{equation}
{\mathsf G}_{{R,L}}=\frac{e^{2}}{h}\sum_{\alpha\in{L}%
}^{^{{L}}M}\sum_{\beta\in{R}}^{^{{R}}M}T_{\beta,\alpha}.
\label{eq_multichannel}%
\end{equation}
This can be written  by combining Eqs. (\ref{eq_Fisher-Lee+DP}) and
(\ref{eq_multichannel})into a compact matrix notation:
\begin{equation}
{\mathsf G}_{{R,L}}=\frac{e^{2}}{h}4\,\,{\rm Tr}\left[
{\mathbf\Gamma}_{{R}}(\varepsilon)\,{\mathbf G}_{{R,L}}
^{R}(\varepsilon){\mathbf\Gamma}_{{L}}(\varepsilon)\,{\mathbf G}
_{{L,R}}^{A}(\varepsilon)\right]. 
\label{eq_TRACE}
\end{equation}
The sum of initial states at left is the result of the product of the
diagonal form of the broadening matrix
${{\mathbf \Gamma}}_{L}(\varepsilon)$, while the trace corresponds to
the sum over final states.

\section{Applications of Landauer's coherent conductance}

\subsection{Lyapunov Exponents and Localization Length.}

The calculation of actual eigenstates in disordered potentials of an
infinite system is a very subtle mathematical problem. In contrast,
the computation of Green's functions and conductances of arbitrary
size can be done at a relatively low computational cost. The sparse
nature of the single particle Hamiltonian when it is expressed in a
localized basis (tight-binding approximation) allows the
implementation of a recursive calculation of the Green's
function. These methods, originally devised to deal with inhomogeneous
and disordered systems, allow the calculation of the quantum
transmittances such as $T_{{R,L}}$. Typically, it requires $L_{x}/a$
operations (products and inversions) with matrices of size $M\times
M$.\bigskip\

Historically, Landauer's reasonings were decisive in the development
of the theory of electronic localization in disordered
systems\cite{z--Pasta-MCF+G,z--MacKinnon-review}. In particular, the
method developed above can be used to proof rigorously that in one
dimension there are no extended states. Every eigenstate should decay
exponentially away of some particular localization center. The exponential 
rate corresponds to the Lyapunov exponent $\gamma$ of the iterative
Dyson Equation. The associated length, $\xi=1/\gamma,$ is called
localization length.

The Lyapunov exponent can be evaluated from the Green's functions $G_{1,n}$ and
$G_{n,n}$ using the limit relations%

\begin{equation}
\gamma(\varepsilon)=\frac{1}{a}\lim_{L\rightarrow\infty}\left[  \frac{1}{L}%
\ln\left|  \frac{G_{L,L}(\varepsilon)}{G_{1,L}(\varepsilon)}\right|  \right]
. \label{eq_def-Lyapunov}%
\end{equation}
This definition is completely general and works either for ordered and
disordered situations. To check how it works, let us first consider the
1-d {\it ordered chain}. The solution of Eq. (\ref{eq_def-Lyapunov}) is
analytic
\begin{equation}
\gamma(\varepsilon)=\left\{
\begin{array}
[c]{c}%
\frac{1}{a}{\rm arccosh}\left[  \displaystyle\frac{\varepsilon-E_{0}}{2V}\right]
\,\,\ \ \,\,{\rm for}\,\,\left|  \varepsilon-E_{0}\right|  >\left|
2V\right|  ,\\
0\,\,\,\,\,\,\,\,\,\ \ \ \ \ \ \ \ \ \ \ \ \,\,\,{\rm for}\,\,\left|
\varepsilon-E_{0}\right|  <\left|  2V\right|  .
\end{array}
\right.  \label{eq_Lyap-1d-chain}
\end{equation}
This clarifies that $\gamma(\varepsilon)$ is just the analytical
continuation of the wave vector $k(\varepsilon).$ Already in this
case, we can identify two regions: one which coincides with the Bloch
band where localization length is infinite and the rest, where the
states, if any, are localized. This is the region where the described
states in the branching region and the surface states lie.

\subsection{Localization in a Conjugated Polymer.}

\begin{figure}[tbp]
\narrowtext
\centering \leavevmode
\center{\epsfig{file=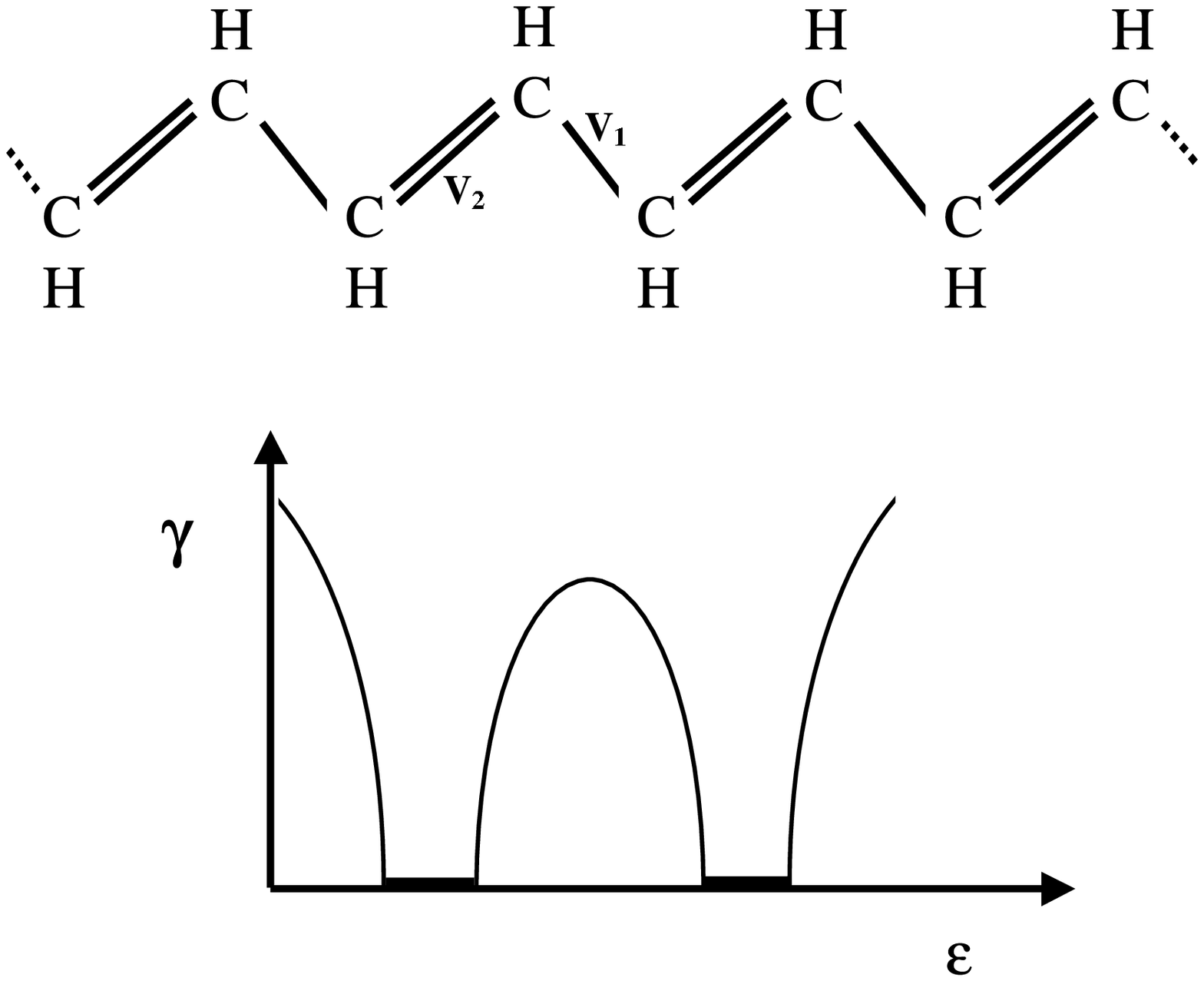,width=7.5cm}} 
\vspace{0.5cm}
\caption{In the upper panel the polyacetilene macromolecule is represented. 
Double bonds and simple bonds alternate. This gives an alternation in
the energies associated to $\pi$-$\pi$ bonding. The lower scheme is
the Lyapunov exponent as a function of the energy. Scales are
discussed in the text.}
\label{fig_LYAPpolymer}
\end{figure}

In order to consider a more realistic example of specific interest in molecular
electronics, let us consider a model \cite{z--SSH} for \ the simplest
conjugated polymer: polyacetilene. This can be described by the
Hamiltonian,
\begin{equation}
\hat{\mathcal{H}}_{0}=\sum_{n({\rm odd})}\left\{  V_{2}\left|  n\right\rangle
\left\langle n+1\right|  +V_{1}\left|  n+1\right\rangle \left\langle
n+2\right|  +{\rm c.c.}\right\}  , 
\label{eq_POLYacetilene}
\end{equation}
where $V_{2}=-(V_{\pi}+\delta)$ and $V_{1}=-(V_{\pi}-\delta)$ where
$V_{\pi }>0$ is the energy associated with a $\pi-\pi$ bonding. This
is represented schematically in Fig.\ref{fig_LYAPpolymer}. $\delta$ is
the additional energy involved in the dimerization to form the
alternate double bond (conjugated system). Hence, the unit cell has
two C-H monomers. Again, the Lyapunov exponent can be calculated
either analytically or numerically giving
\wide{m}{
\begin{equation}
\gamma(\varepsilon)=\left\{
\begin{array}
[c]{c}  \frac{1}{a} {\rm arccosh}\left[  \frac{\varepsilon^{2}-(V_{1}
^{\,\,\,2}+V_{2}^{\,\,\,2})}{2V_{2}^{{}}V_{1}^{\,\,\,}}\right]
{\rm  at \,\,the\,\,gaps}\\
 0  {\rm  at \,\,the\,\,valence\,band,}V_{2}+V_{1}<\varepsilon
<V_{2}-V_{1}\\
 0  {\rm  at \,\,the\,\,conduction\,band,}-V_{2}+V_{1}<\varepsilon<-V_{2}-V_{1}.
\end{array}\right.  \label{eq_LYAPolyacetilene}%
\end{equation}
}

This simple situation hints at the solution of a the more complex disordered
systems. Conceptually, a disordered sample can be visualized as a periodic
array of disordered unit cells of size $L=Na$, with a progressively increasing
$N$. Then, the localization transition appears as the dominance of the
``gaps'' over the ``bands'' whose width (spectral support) decreases as
$V\exp[-\gamma Na]$. With the decimation methodology developed above, the
numerical evaluation of the Lyapunov exponent is straightforward.

\subsection{Metal-Insulator Transition.}

In one dimensional systems, the exponential decrease of the
conductance follows directly from the statistical properties of the
transmission probability $T\sim\exp\left[ -L_{x}/\xi\right] $ in a
disordered system. Therefore,
the \ `four-probe ' Landauer's conductance given by
Eq. (\ref{eq_4p-Landauer}), by describing the intrinsic properties of
the ``sample'', contains the correct scaling behavior of the
conductance ranging from the non-extensive behavior when
$L_{x}\gg\xi,$%
\begin{equation}
{\mathsf G}_{{R,L}}^{{\rm f.p.}}(L_{x})\simeq\frac{e^{2}}{h}%
\exp\left[  -L_{x}/\xi\right]  , \label{eq_conductance-localized}%
\end{equation}
to the expected Ohmic behavior described by Drude's law in terms of the mean
free path $\ell=\xi/2,$
\begin{equation}
{\mathsf G}_{{R},{L}}^{{\rm f.p.}}(L_{x})=e^{2}N_{0}%
\,v_{F}\,\ell/L_{x}, \label{eq_Drude}%
\end{equation}
when $L_{x}<\ell.$ Statistical subtleties aside, this last result is
obtained by expanding the transmittance in its lowest order in $L_{x}/\xi$.

\begin{figure}[tbp]
\narrowtext
\centering \leavevmode
\center{\epsfig{file=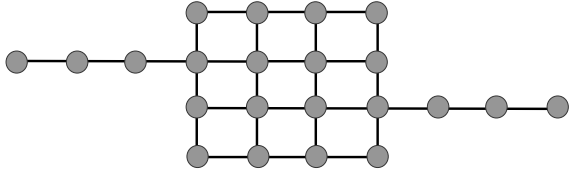,width=7.5cm}} 
\vspace{0.5cm}
\caption{Scheme of the use of the calculation of 1-d transmittance to obtain the
transport properties of a 2-d system.}
\label{fig_square+2leads}
\end{figure}

\begin{figure}[tbp]
\narrowtext
\centering \leavevmode
\center{\epsfig{file=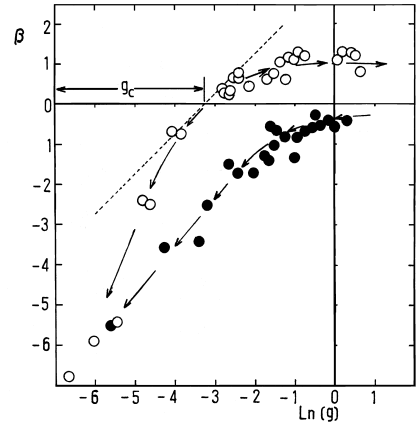,width=5.5cm}} 
\vspace{0.5cm}
\caption{Numerical scaling function $\beta=(L/g_{L})$ $\left(  
\Delta g_{L}/\Delta L\right) $ evaluated from the adimensional conductance 
$g_{L}$ of samples of 2-$d$ and 3-$d$ system size $L^{d}.$ We use the
Anderson model at $\varepsilon_{F}=0$. Data for disorder
$W/V=4,5,6,7,8,10,14$ and sizes up to $L^{2}=20\times20a^{2}$ collapse
in the 2-$d$ curve of the filled circles.  Data for disorder
$W/V=6,8,10,14,15.5,16.5,17$ and $18$ \ and sizes up to
$L^{3}=6\times6\times6a^{3}$ collapse in the 3-$d$ curve of the empty
circles.  In 3-d the critical conductance is $g_{c}\simeq3.2$ and
corresponds to a critical disorder $W_{c}/V\simeq16$.}
\label{fig_num-scaling-curve}
\end{figure}

A similar reasoning can be applied to higher dimensional systems
($d\geq2$) with hyper-cubic shape. The conductance can be calculated
with Eq.(\ref{eq_4p-Landauer}) by attaching two one-dimensional
leads. The metal-insulator transition is possible because the curve
$\xi(L)=2\ell (L/a)^{d-1},$ valid for weak scattering, bends down to
an asymptotic value $\xi(L)\rightarrow\xi_{\infty}$.  This effect
manifests itself in a universal behavior. One considers the
conductance of samples of size $L^{d}=\left( Na\right)^{d}$ with
Anderson disorder with two one dimensional leads attached as shown in
Fig.\ref{fig_square+2leads}. The adimensional conductance $g_{L}=\frac
{h}{e^{2}}{\mathsf G}_{{R,L}}^{{\rm f.p.}}(L)$ is calculated using
Eq.(\ref{eq_4p-Landauer}). For each dimensionality, all sizes and
disorders strength scale into a single scaling curve
\begin{equation}
\beta_{d}=\frac{L}{g_{L}}\frac{{\rm d}g_{L}}{{\rm d}L}.
\label{eq_Beta-function}%
\end{equation}
This is shown in Fig.\ref{fig_num-scaling-curve} for dimensions $d=2$ and
$d=3$ as a function of $\ln[g_{L}].$%

The arrows indicate the directions in which $g_{L}$ moves as the size
of the system increases. The fact that $\beta_{d=2}$ (full circles)
flows from $\beta_{2}=0$ towards $-\infty$ indicates that, for any
positive disorder all states are localized. In contrast, in a
$\beta_{d=3}$ the curve can flow either towards $\beta_{2}=1,$ if the
disorder is weak enough ($W<W_{c}$), or towards
$\beta_{2}\rightarrow-\infty$ for disorder above certain critical
value ($W>W_{c}$ \ with $W_{c}\simeq6.5V$). This shows that the MIT is
a critical phenomenon as any other thermodynamical phase transition.

The point we want to stress here is that the ``non-invasive voltage
probes'' of the Landauer picture rescue a ``local'' meaning for the
conductance and emphasizes the non-extensive behavior introduced by
quantum interferences. The presence of decoherent processes or
voltmeters on a scale $L_{\phi}$ ``break'' the quantum conductance
into an incoherent sum of $\frac{L_{\phi}}{L_{x}}$ separate pieces
where the transport is coherent and hence described by
Eq. (\ref{eq_4p-Landauer}). The total conductance becomes ${\mathsf
G}_{{R,L}}(L_{x}) =\frac{L_{\phi}}{L_{x}}{\mathsf G}_{{R,L}}^{{\rm
f.p.}}(L_{\phi})$ recovering the extensive Ohmic behavior.

\subsection{The Aharonov-Bohm effect and non-local properties.}

Setting the Zeeman effect aside, the main consequence of a magnetic field is
to affect the wave function's phase in
\[
\phi=\frac{e}{hc}\int\vec{A}\cdot {\mathrm d}\vec{l}.
\]
In the Peierls substitution, appropriate for discrete models, this is
achieved by modifying the coupling $V\mapsto V\exp[{\rm i}\phi].$

\begin{figure}[tbp]
\narrowtext
\centering \leavevmode
\center{\epsfig{file=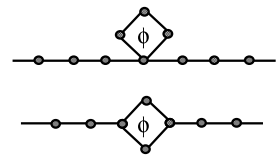,width=3.5cm}} 
\vspace{0.5cm}
\caption{Two tight binding representations of path loops affecting the
conductance in a non-local fashion as discussed in the text.}
\label{fig_loops}
\end{figure}

When we have a system \cite{z--Webb-AB} as that shown in Fig.\ref{fig_loops}, the
application of the magnetic field will result in an oscillation of the
conductance\cite{z--AAS-effect} due to the interference of the probability amplitudes
propagated through different branches of the system.

Let's assume that in absence of field these pathways superpose in the form
of maximally destructive interference in the forward direction and have
constructive probability of return $P_{0,0}$ (this is localization)
\begin{equation}
P_{0,0}=\left|  \sum\limits_{i({\rm paths})}u_{0}^{i}\right|  ^{2}%
=\sum\limits_{i}\left|  u_{0}^{i}\right|  ^{2}+\sum\limits_{i\neq j}u_{0}%
^{i}u_{0}^{j\ast} \label{eq_pathways}%
\end{equation}

The first term in the right side is the contribution of uncorrelated
pathways.  The second term gives the interference effects, it is
affected by the magnetic field and contains the non-local effects
intrinsic to the Schr\"{o}dinger equation.
In presence of a gauge field $A,$ destructive interference is modified
producing an increase of the transmittance. This is a positive
magnetoconductance\cite{z--Medina-Pastawski}, a characteristic effect of coherent systems at
very low fields. If the inelastic events along the pathways start
contributing to the loss of phase memory, the field effect becomes
progressively weaker until it eventually disappears.  As we will see
bellow, a very strong magnetic field causes interferences in the short
distances (toward the Landau levels) therefore giving a new tendency
toward localization.

\subsection{Conductance quantization in 2-d}

\begin{figure}[tbp]
\narrowtext
\centering \leavevmode
\center{\epsfig{file=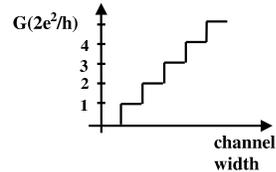,width=3.5cm}} 
\vspace{0.5cm}
\caption{Quantized conductance as a function of channel width.}
\label{fig_stepsG}
\end{figure}

Extending what is observed in a 1-$d$ ordered chain, where
transmittances are either 0 or 1 depending on the energy of the
injected excitation; when one considers the 1$^{+}$-$d$ (strips) and
real 2-$d$ systems, we have to account for the different propagating
modes. As the available energy increases, new lateral modes becomes
available increasing the total current. According to
Eq. (\ref{eq_multichannel}) the conductance is bounded by

\begin{equation}
{\mathsf G}_{{R,L}}\,\leq\frac{e^{2}}{h}\min[\,^{{L}}M_{\left(
\varepsilon_{F}\right)  },\,^{{R}}M_{\left(  \varepsilon_{F}\right)
}]. \label{eq_multichannel G}%
\end{equation}
For perfect transmitting samples, a situation that requires order and
reflection symmetry ($^{{L}}M_{\left( \varepsilon_{F}\right)
}=\,^{{R}}M_{\left( \varepsilon_{F}\right) })$, $T_{j,i}$ is
either $1$ or $0$ and one obtains the conductance quantization in
integer multiples of $e^{2}/h$. This effect is spectacularly verified
in specially tailored nano-structures. The conductance's discrete
steps are shown in Fig.\ref{fig_stepsG}.

\begin{figure}[tbp]
\narrowtext
\centering \leavevmode
\center{\epsfig{file=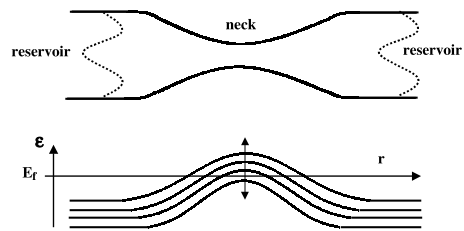,width=6.0cm}} 
\vspace{0.5cm}
\caption{Geometry of neck formed when a nanowire is pulled lengthwise. As
the transverse dimension shrinks conductance channels are excluded as shown
in the lower panel where we show a  scheme of
the transversal modes energies as a function of position.}
\label{fig_contact+energy} 
\end{figure}

Such an effect is restricted to artificial devices\cite{Wharam}. The
same effect occurs when there is a metallic bridge between two
electrodes and these are pulled apart. The bridge stretches becoming
progressively thinner until its breakdown. This process is monitored
by the carriers, which, as the bridge becomes narrower, loose
propagating channels below the Fermi energy. As a consequence the
conductance shows clear steps such as those shown in in
Fig.\ref{fig_stepsG}. The adiabatic approximation, justified by the
high electronic speeds as compared to any modification of the contact
structure, allows the use of a simple model where the time is
parametrically associated with the width represented in
Fig.\ref{fig_contact+energy}.

\subsection{Hall Effect}

\begin{figure}[tbp]
\narrowtext
\centering \leavevmode
\center{\epsfig{file=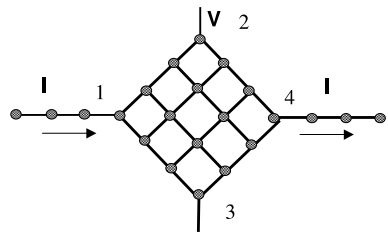,width=5.0cm}} 
\vspace{0.5cm}
\caption{Geometrical representation of the model used to evaluate the Hall 
effect.}
\label{fig_red-Hall}
\end{figure}

In the classic theory of the Hall effect the external magnetic field
produces a coefficient $R_{H}=-1/ne$ which is negative if the carriers
are electrons and is inversely proportional to the electron
density. In a two dimensional system, we can evaluate the Hall
effect at low fields by using a simple model that has two 1-$d$
contacts to inject and measure the current and two additional ones to
measure the Hall voltage. These are shown in Fig.\ref{fig_red-Hall}

Since the chemical potentials and voltages are related by ${\mathsf V}_{p}
=\mu_{p}/e,$ Eq.(\ref{eq_current-multilead}) becomes
\[
{\mathsf I}_{p}=\sum_{q=1}^{4}[{\mathsf G}_{q,p}{\mathsf V}_{p}-{\mathsf G}
_{p,q}{\mathsf V}_{q}].
\]
When all the chemical potentials are equal, there is no current and
$\sum\limits_{q}{\mathsf G}_{p,q}=\sum\limits_{p}{\mathsf G}_{q,p}$ from which
\[
{{\mathsf I}}_{p}=\sum\limits_{q=1}^{4}{\mathsf G}_{p,q}({\mathsf V}_{p}
-{\mathsf V}_{q}).
\]
We can always choose one of the voltages, e.g. one of the current leads as a
reference setting ${\mathsf V}_{4}=0.$ Therefore, the problem simplifies to three linear
equations
\wide{m}{
\begin{eqnarray*}
\left(
\begin{array}
[c]{c}
{\mathsf I}_{1}\\
{\mathsf I}_{2}\\
{\mathsf I}_{3}
\end{array}
\right)   &  = &\left(
\begin{array}
[c]{ccc}
{\mathsf G}_{2,1}+{\mathsf G}_{3,1}+{\mathsf G}_{4,1} & -{\mathsf G}_{1,2} &
-{\mathsf G}_{1,3}\\
-{\mathsf G}_{2,1} & {\mathsf G}_{1,2}+{\mathsf G}_{3,2}+{\mathsf G}_{4,2} &
-{\mathsf G}_{2,3}\\
-{\mathsf G}_{3,1} & -{\mathsf G}_{3,2} & {\mathsf G}_{1,3}+{\mathsf G}
_{2,3}+{\mathsf G}_{4,3}
\end{array}
\right)  \left(
\begin{array}
[c]{c}
{\mathsf V}_{1}\\
{\mathsf V}_{2}\\
{\mathsf V}_{3}
\end{array}
\right). \\
\end{eqnarray*}
In short ${\mathsf I}_{i}    = \sum_{j}M_{i,j}{\mathsf V}_{j}$, where we invert the matrix of conductances $\mathbf M$ obtaining
\[
\left(
\begin{array}
[c]{c}
{\mathsf V}_{1}\\
{\mathsf V}_{2}\\
{\mathsf V}_{3}%
\end{array}
\right)  =\left(
\begin{array}
[c]{ccc}%
{\mathsf R}_{1,1} & {\mathsf R}_{1,2} & {\mathsf R}_{1,3}\\
{\mathsf R}_{2,1} & {\mathsf R}_{2,2} & {\mathsf R}_{2,3}\\
{\mathsf R}_{3,1} & {\mathsf R}_{3,2} & {\mathsf R}_{3,3}%
\end{array}
\right)  \left(
\begin{array}
[c]{c}%
{\mathsf I}_{1}\\
{\mathsf I}_{2}\\
{\mathsf I}_{3}%
\end{array}
\right)  .
\]
Let's see one of its elements,

\[
{\mathsf R}_{1,1}=\frac{({\mathsf G}_{2,3}+{\mathsf G}_{2,4}+{\mathsf G}
_{2,1})({\mathsf G}_{3,1}+{\mathsf G}_{3,2}+{\mathsf G}_{3,4})-{\mathsf G}
_{2,3}{\mathsf G}_{3,2}}{\det[{\mathbf M}]}.
\]
}
The voltage between 2 and 3 is related to the total current%

\[
{\mathsf R_H}=\frac{{\mathsf V}}{{\mathsf I}}=\frac{{\mathsf V}_{2}-{\mathsf V}_{3}
}{{\mathsf I}_{1}}={\mathsf R}_{2,1}-{\mathsf R}_{3,1}
\]
which will be zero in absence of a magnetic field. Otherwise, the
dependence on magnetic field implicit in the quantum mechanical evaluation of
${\mathsf R}_{2,1}$ and ${\mathsf R}_{3,1},$ gives the Hall resistance.

There are certain symmetry relations that should be obeyed on the basis of the
so called Onsager-Casimir relations; in the linear response regime the Hall
resistance obeys 
\begin{equation}
{\mathsf R}_H(B)=-{\mathsf R}_H(-B),
\end{equation}
where $B$ is the applied field. The previous relation involves the minus
sign because the Hall voltage is reversed in the field. Nevertheless, if one
simultaneouly exchanges the voltage probes there is no change. On the other
hand the ordinary longitudinal resistance does not change under the reversal of
the field. The symmetry relation can be summarized as
\begin{equation}
{\mathsf R}_{xx}(B)={\mathsf R}_{xx}(-B),
\end{equation}
and
\begin{equation}
{\mathsf R}_{xy}(B)={\mathsf R}_{yx}(-B).
\end{equation}
These symmetries are well satisfied for an homogeneous sample. On the other
hand, for inhomogeneous samples no specific symmetry is experimentally found upon
reversing the field. This is due to irregular current patterns occuring in the
sampple mixing both $xx$ and $xy$ components the first one being symmetric, and the
second antisymmetric, in the field. A more general relatioship for a four probe
device can be given as follows: if one defines the resistance ${\mathsf R}_{mn,kl}$
as that taken when measuring the voltage between probes $k,l$ while putting current
$I$ between probes $m,n$ (imput at $m$ output at $n$)
\begin{equation}
{\mathsf R}_{mn,kl}=\frac{{\mathsf V_k-V_l}}{I}.
\end{equation}
The {\it reciprocity relations} are then expressed as
\begin{equation}
{\mathsf R}_{kl,mn}(B)={\mathsf R}_{mn,kl}(-B).
\end{equation} 

The Integer Quantum Hall Effect IQHE\cite{z--Büttiker-Hall} follows
from the use of Eq. \ref{eq_current-multilead} in the condition that
strong magnetic fields confine the propagating electrons to the sample
boundaries. The appearance of these new spatial selection rules
precludes backward scattering and
decoherence\cite{z--Q-Hall+dephasing}, yielding the unitary
transmittances responsible for the IQHE.

\section{Decoherence in Quantum Transport}

\subsection{Phenomenology}\label{Phenomenology}

\begin{figure}[tbp]
\narrowtext
\centering \leavevmode
\center{\epsfig{file=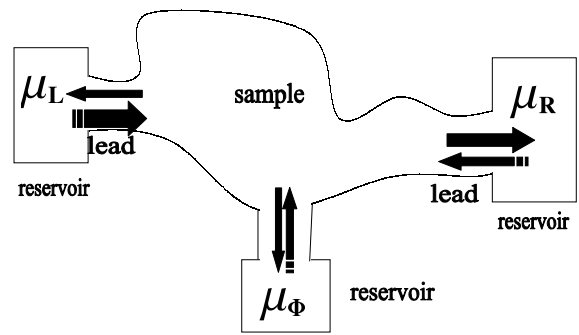,width=7.5cm}} 
\vspace{0.3cm}
\caption{Representation of a three probe measurement. The volt-meter may be 
strongly coupled and is a source of decoherence.}
\label{fig_3reservorios}
\end{figure}

It is obvious that because of the experimental limits for the coherent
description introduced above one should observe \cite{z--Webb-decoh}
important departures from those predictions. A first alternative to
include decoherence in quantum transport was inspired by the
Landauer's formulation. There, the leads, while accepting a quantum
description of their spectra and their ability to propagate
excitations, are the ultimate source of irreversibility and
decoherence: Electrons leaving the leads toward the sample are
completely incoherent with the electrons coming from the other leads
(see Fig. \ref{fig_3reservorios}). In fact, it is obvious that a wire
connected to a voltmeter, by ``measuring'' the number of electrons in
it, must produce some form of collapse of the wave function leading to
decoherence.

\begin{figure}[tbp]
\narrowtext
\centering \leavevmode
\center{\epsfig{file=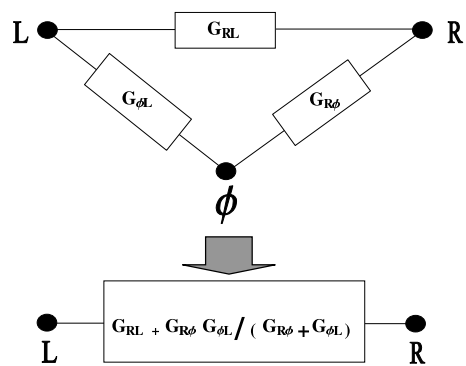,width=6.0cm}} 
\vspace{0.2cm}
\caption{Classical circuit representation of the non-classical system 
with quantum coherent and incoherent transport. The coherent
component is a direct resistance between left ($L$) and right ($R$)
electrodes. Quantum mechanics makes these effective resistances
interdependent.}
\label{fig_resistencias}  
\end{figure}

Besides, no net current flows toward a voltmeter. Leads are then a
natural source of decoherence which can be readily described in the
Landauer's picture if one uses the Landauer conductances together with
the Kirchhoff balance equations. This fact was firstly realized by
M. B\"{u}ttiker\cite{z--Büttiker-decoh}. The procedure, resulting from
the application of Kirchhoff law to each lead, is called the
Landauer-B\"{u}ttiker equation\cite{z--Büttiker-Kirchhoff}. Let us see
how it works for the case of a single voltmeter. In matrix form
\wide{m}{
\begin{equation}
\left(
\begin{array}
[c]{l}
{\mathsf I}_{{L}}\\
{\mathsf I}_{{\phi}}\\
{\mathsf I}_{{R}}
\end{array}
\right)  =\left(
\begin{array}
[c]{lll}%
-\left[ {\mathsf G}_{{R,L}}+{\mathsf G}_{{\phi,L}}\right]  &
\,\,\,\ \,\,\,\,\,\,\,\,\ \,{\mathsf G}_{{L,\phi}} &
\,\,\,\ \,\,\,\,\,\,\,\,{\mathsf G}_{{L,R}}\\
\,\,\,\ \,\,\,\,\,\,\,\,\ \,{\mathsf G}_{{\phi,L}} & -\left[
{\mathsf G}_{{R,\phi}}+{\mathsf G}_{{L,\phi}}\right]  &
\,\,\,\ \,\,\,\,\,\,\,\,\,{\mathsf G}_{{\phi,R}}\\
\,\,\,\ \,\,\,\,\,\,\,\,\ \,{\mathsf G}_{{R,L}} &
\,\,\,\ \,\,\,\,\,\,\,\,\ \,{\mathsf G}_{{R,\phi}} & -\left[
{\mathsf G}_{{\phi,R}}+{\mathsf G}_{{L,R}}\right )
\end{array}
\right]  \left(
\begin{array}
[c]{l}%
{\mathsf V}_{{L}}\\
{\mathsf V}_{{\phi}}\\
{\mathsf V}_{{R}}%
\end{array}
\right). 
\label{eq_linear-currents}
\end{equation}
}
Here, the unknowns are ${\mathsf I}_{{L}},{\mathsf I}_{{R}}$ and
${\mathsf V}_{{\phi}}=\delta\mu_{{\phi}}/e.$ The second equation
must be solved with the voltmeter condition ${I}_{\phi}\equiv0$
\begin{equation}
0=\frac{e}{h}T_{{\phi,L}}(\delta\mu_{{\phi}}-\delta
\mu_{{L}})+\frac{e}{h}T_{{R,\phi}}(\delta\mu_{{\phi}%
}-\delta\mu_{{R}}), \label{eq_Voltmeter-balance}%
\end{equation}
yielding us $\delta\mu_{{\phi}}$ to be introduced in the third
equation
\begin{equation}
{\mathsf I}_{{R}}=\frac{e}{h}T_{{R,L}}(\delta\mu_{{L}%
}-\delta\mu_{{R}})+\frac{e}{h}T_{{R,\phi}}(\delta\mu
_{{L}}-\delta\mu_{{\phi}})
\end{equation}
to obtain the current at the voltage source
\begin{equation}
{\mathsf I}_{{R}}=\frac{e}{h}\widetilde{T}_{{R,L}}(\delta
\mu_{{L}}-\delta\mu_{{R}}) \label{eq_total-current}%
\end{equation}
with
\begin{equation}
\widetilde{T}_{{R,L}}=T_{{R,L}}+\frac{T_{{R,\phi}%
}T_{{\phi},{L}}}{T_{{R,\phi}}+T_{{\phi}%
,{L}}}. \label{eq_incoherent-transmittance}%
\end{equation}
The first term can be identified with the coherent part, while the second is
the incoherent or sequential, i.e. the contribution to the current originated
from particles coming from the voltmeter. This corresponds to an effective
conductance of
\begin{equation}
\widetilde{{\mathsf G}}_{{R,L}}={\mathsf G}_{{R,L}}+({\mathsf G}
_{{R,\phi}}^{\,\,\,\,\,\,\,\,\,\,-1}+{\mathsf G}_{{\phi,L}
}^{\,\,\,\,\,\,\,\,\,\,-1})^{-1}, 
\label{eq_paralel-resistances}
\end{equation}
which can be identified with the electrical circuit of Fig.
\ref{fig_resistencias}. This classical view clarifies the
competition between coherent and incoherent transport. What this
circuit does not hint at is that in Quantum Mechanics one cannot
modify one of the resistances without deeply altering the others.

So far with the phenomenology. The next important step is to connect
these quantities with actual model Hamiltonians. This connection was
made explicit by the contribution of D'Amato and
Pastawski\cite{z--D'Amato-Pastawski}(DP).

\subsection{The D'Amato-Pastawski model for decoherence.}

The DP model refers to a simple way to account for
the infinite degrees of freedom of the thermal bath or the electron
reservoirs.  This follows from our general approach: to obtain exact
solutions to simple problems instead of finding approximate solutions
of complex problems. Let us first review the basic mathematical background that
made possible the selection of a simple Hamiltonian that best represents the complex
sample-environment system. The objective was to use its exact solution in the
Landauer's transport equation. Here, we describe the DP model 
for decoherence and
show how it applies to a simple resonant tunneling system. Consider the
sample's Hamiltonian
\begin{equation}
{\hat{\mathcal H}}_0=\sum_{i=1}^{N}\left\{  E_{i}\left|  i\right\rangle
\left\langle i\right|  +\sum_{j=1\left(  \neq i\right)  }^{N}V_{i,j}\left[
\left|  i\right\rangle \left\langle j\right|  +\left|  j\right\rangle
\left\langle i\right|  \right]  \right\}  , \label{eq_Ho-chain}%
\end{equation}
where \thinspace$i$ and $j$ indicate sites on a lattice. Notice that
interactions are not restricted to nearest neighbors. However, for the
usual short range interactions, the Hamiltonian matrix has the
advantage of being sparse. The local dephasing field is represented by
\begin{equation}
^{{\phi}}\hat{\Sigma}^{R}=\sum_{i}^{N}-({\rm i\,\,}) ^{{\phi}
}\Gamma\,\,\left|  i\right\rangle \left\langle i\right|,
\label{eq_Sigma-DP model}
\end{equation}
where $^{{\phi}}\Gamma=\hbar/(2\tau_{\phi})$ and we consider for
simplicity only two one-dimensional current leads ${L}$ and
${R}$ connected at sites 1 and $N$ respectively
\begin{equation}
^{{\rm leads}}\hat{\Sigma}^{R}=-{\rm i}\left(  \,^{{L}}
\Gamma\,\,\left|  1\right\rangle \left\langle 1\right|  +\,^{{R}}
\Gamma\,\,\left|  N\right\rangle \left\langle N\right|  \right).
\label{eq_Sigma-escapes}
\end{equation}
We see that the $1^{{\rm st}}$ site has escape contributions towards
\textit{both}, the current lead at the left, $^{{L}}\Gamma_{1},$
\textit{and} the inelastic channel associated to this site, $^{{\phi}%
}\Gamma_{1}.$  The on-site chemical potential will ensure
that no net current flows through the latter channel.

\subsection{The solution for incoherent transport}

The adimensional conductances are just the transmission probabilities. In
principle, they can be computed in terms of the Green's function according to
Eq. (\ref{eq_Fisher-Lee+DP}). To simplify the notation we define the total
transmission from each site as
\begin{equation}
\left(1/g_{i}\right)  \equiv\sum_{j=0}^{N+1}T_{j,i}=\left\{
\begin{array}
[c]{c}
4\pi N_{1}^{{}}\,^{{L}}\Gamma_{1}\,\,\,\,\,\,\,\,{\rm for}%
\,\,i=0\,\,\,\,\,\ \ \ \,\,\\
4\pi N_{i\,}^{{}}\,^{{\phi}}\Gamma_{i}\,\,\,\,\,\,{\,\rm for}%
\,\,1\leq i\leq N\\
4\pi N_{N}^{{}}\,^{{R}}\Gamma_{N}\,\,\,\,\,{\rm for}\,\,i=N+1
\end{array}
\right. 
\label{eq_total-escape}
\end{equation}
The last equality follows from the optical theorem. The balance equation
becomes
\begin{equation}
{\mathsf I}_{i}\equiv0=-\left(  1/g_{i}\right)  \delta\mu_{i}+\sum_{j=0}%
^{N+1}T_{i,j}\delta\mu_{j}, 
\label{eq_localcurrents}
\end{equation}
where the sum adds all the electrons that emerge from a last collision
at other sites ($j$'s) and propagate coherently to site $i.$ These
include the electrons coming from the current source
i.e. $T_{i,{L}}\delta
\mu_{{L}}$ and the current drain. However, since we refer all voltages
to the last one, $T_{i,{R}}\delta\mu_{{R}.}\equiv0$. We remark that
here we do not exclude the $i$-th site from the summations in Eqs.
(\ref{eq_localcurrents}) and (\ref{eq_total-escape})$.$ While this has
no consequences in the steady state, they become relevant in the time
dependent formulation. To fix the physical interpretation we emphasize
that the last term accounts for the electrons that emerging from a
dephasing collision at site $j$ will propagate coherently to site $i$
where they have a dephasing collision. The first term accounts for all
the electrons that emerge from this collision on site $i$ to have a
further dephasing collision either in the sample or in the leads. The
net current is identically zero at any dephasing\ \ channel
(``lead''). The other two equations are
\begin{eqnarray}
{\mathsf I}_{{L}}  &  \equiv & -{\mathsf I}=-\left(  1/g_{i}\right)
\delta\mu_{{L}}+\sum_{j=0}^{N}T_{i,j}\delta\mu_{j}
,\label{eq_currentleads}\\
{\mathsf I}_{{R}}  &  \equiv & {\mathsf I}=-\left(  1/g_{i}\right)
\delta\mu_{i}+\sum_{j=0}^{N+1}T_{i,j}\delta\mu_{j}.\nonumber
\end{eqnarray}
Here, we need the local chemical potentials. They can be obtained from Eq.
(\ref{eq_localcurrents}). In a compact notation, these coefficients can be
arranged in a matrix form which excludes the leads that are current source and
sink
\wide{t}{
\begin{equation}
\mathbf{W}=\left (
\begin{array}
[c]{lllll}%
T_{1,1}-1/g_{1} & T_{1,2} & T_{1,3} & \cdots &  T_{1,N}\\
T_{2,1} & T_{2,2}-1/g_{2} & T_{2,3} & \cdots &  T_{2,N}\\
T_{3,1} & T_{3,2} & T_{3,3}-1/g_{3} & \cdots &  T_{3,N}\\
\vdots & \vdots & \vdots & \ddots & \vdots\\
T_{N,1} & T_{N,2} & T_{N,3} & \cdots &  T_{N,N}-1/g_{N}%
\end{array}
\right )  , \label{eq_W-matrix}%
\end{equation}
}
from which the chemical potential in each site can be calculated as

\begin{equation}
\delta\mu_{i}=\sum_{j=1}^{N}\left[  \mathbf{W}^{-1}\right]  _{i,j}%
T_{j,0}\,\delta\mu_{0}. \label{eq_local-mu}%
\end{equation}
Replacing these chemical potentials back in Eq. (\ref{eq_localcurrents}) the
effective transmission can be calculated
\begin{equation}%
\begin{array}
[c]{ccc}%
\widetilde{T}_{{R,L}}= & \underbrace{T_{{R,L}}}+ &
\underbrace{\sum_{j=1}^{N}\sum_{i=1}^{N}T_{{R},j}\left[  \mathbf{W}%
^{-1}\right]  _{j,i}T_{i,{L}}}.\\
& {\rm coherent} & {\rm incoherent}%
\end{array}
\label{eq_total transmision}%
\end{equation}
The right side contains two contributions: the first comes from electrons that
propagate quantum \textit{coherently} through the sample, the second contains
the \textit{incoherent} contributions due to electrons that suffer their first
collision at site $i$ and their last at site $j$.

\begin{figure}[tbp]
\narrowtext
\centering \leavevmode
\center{\epsfig{file=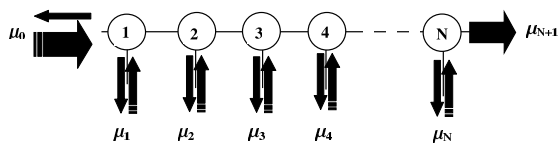,width=7.5cm}} 
\vspace{0.5cm}
\caption{Pictorial representation of the D'Amato-Pastawski model for 
the case of a linear chain.}
\label{fig_Damato}
\end{figure}

Until now the procedure has been completely general, there is no
assumption involving the dimensionality or geometry of the sample. The
system of Fig.\ref{fig_Damato} was adopted by DP only because it
has a simple analytical solution for various situations ranging from
tunneling to ballistic transport.  We summarize the procedure for the
linear response calculation. First, we calculated the complete Green's
function in a tight binding model. With the Green's functions we then
evaluate the transmittances between every pair of sites\ in the sample
(i.e. nodes in the discrete equation) and write the transmittance
matrix $\mathbf{W}.$ Then, we solve for the current conservation
equations that involves the inversion of $\mathbf{W}$.

What are the limitations of this model? A conceptual one is the
momentum demolition produced by the localized scattering
model. Therefore, we go directly from quantum ballistic description to
a classical diffusive one. To describe the transition from quantum
ballistic to classical ballistic, one should modify the model to have
the scattering defined in momentum or energy basis.

The other aspect is merely computational. Since the resulting matrix
$\mathbf{W} $ is no longer sparse, this inversion is done at the full
computational cost. A physically appealing alternative to matrix
inversion was proposed in DP. The idea was to expand the inverse
matrix in series on the dephasing collisions, resulting in
\begin{eqnarray}
\widetilde{T}_{{R,L}}  &  = &T_{{R,L}}+\sum_{i}T_{{R},i}
g_{i}T_{i,{L}}+\sum_{i}\sum_{j}T_{{R},i}g_{i}T_{i,j}
g_{j}T_{i,{L}}\nonumber\\
&  + &\sum_{i}\sum_{j}\sum_{l}T_{{R},i}g_{i}T_{i,j}g_{j}T_{j,l}
g_{l}T_{l,{L}}+\ldots\label{eq_T-series}
\end{eqnarray}
The formal equivalence with the self-energy expansion of Matsubara and
Toyosawa in terms of locators\cite{z--locator} or local Green's
function justifies identifying $g_{i}$ as a \textit{locator }for the
classical Markovian equation for the density\cite{z--GLBE1} generated
by the transition probabilities.

As an example of use of Eq. (\ref{eq_T-series}), one can readily apply
it for the model of Section\ref{Phenomenology} obtaining
Eq.(\ref{eq_incoherent-transmittance}).  This also constitutes the
basis for a perturbative method of calculating the conductance with a
substantially reduced computational cost. This strategy has been applied with considerable success to explain the stability of the
Quantum Hall Effect\cite{z--Q-Hall+dephasing} against scattering and
dephasing.

\begin{figure}[tbp]
\narrowtext
\centering \leavevmode
\center{\epsfig{file=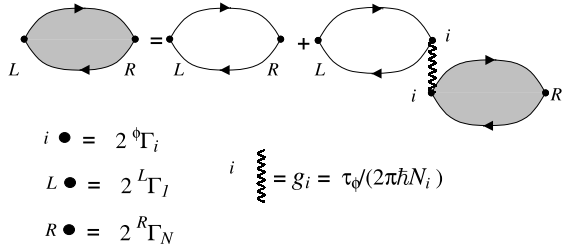,width=7.5cm}} 
\vspace{0.5cm}
\caption{Feynman Diagram for the Dyson Equation of the transmittance. 
It is equivalent to a particle-hole Green's function in the ladder
approximation where the usual interaction rung connecting the two propagators is here represented by a dot.}
\label{fig_Feynman-T}
\end{figure}

Notice that Eq. (\ref{eq_T-series}) can also be rearranged as
\begin{equation}
\begin{array}
[c]{ccccc}
\underbrace{\widetilde{T}_{{R,L}}} & =\underbrace{T_{{R,L}}
}\,\,\,+ & \displaystyle\sum_{i=1}^{N}\underbrace{
\widetilde{T}_{{R},i}} & \times\,\,\,\,\,\ \underbrace{g_{i}}\,\,\,\,\times
& \underbrace{T_{i,{L}}}\\
\mathrm{total} & =\mathrm{coherent} & +\,\,\,\mathrm{total}\times &
\mathrm{1}^{st}\mathrm{\,collision}\times & \mathrm{coherent}
\end{array}
\label{eq_T-Dyson}%
\end{equation}
The summation on the right hand side has the structure of the Dyson
equation. It is a Bethe-Salpeter equation. This is graphically
represented in Fig.\ref{fig_Feynman-T}. We notice that according to
the optical theorem $g_{i}=2\pi\hbar\tau_{\phi} N_{i},$
while both transmittances entering the vertices in the figure are
proportional to $1/\tau_{\phi}$ \ the whole product involved in the
vertex is proportional to the dephasing rate. The arrows make explicit
that transmittances are the product of a retarded (electron) and an
advanced (hole) Green's function. To describe diffusive scattering by
impurities one must replace the whole electron-hole pair by its
ensemble average calculated in the ladder approximation, or eventually
include their quantum corrections (see Ref. \cite{z--GLBE1}).

Since the transmittance is essentially a two-particle Green's
function, now it is a particular case of the Bethe-Salpeter equation
which is solved in the ``ladder'' approximation which for the present
model is exact (see appendix B in Ref. \cite{z--GLBE1} for a detailed
formalization of this picture with the Keldysh formalism). In summary,
depending on the approximation used for the density propagator, the
self-consistent solution contains the metallic transport, the
weak-localization corrections and even the thermally activated hopping
regime. It also can be viewed as a self-consistent Born approximation
\cite{z--Hershfield} for the electron-phonon interaction. One might also want
to introduce other forms of interaction, such as those which conserve
momentum\cite{z--k-conserving,Brouwer}, but we will not extend on this
point here.

Many of the results contained in the D'Amato-Pastawski paper for
ordered and disordered systems were later extended in great detail in
a series of papers by S. Datta and are described in his didactic book
\cite{z--Datta}. In the next section, we will illustrate how the
previous ideas work by considering again our reference toy model for
resonant tunneling.

\subsection{Effects of decoherence in resonant tunneling}

If we consider the ``sample'' to consist of a single
state\cite{z--GLBE2}. If we choose to absorb the energy shifts into
the site energies $\widetilde{E}_{0} =E_{0}+\Delta_{0}\left(
\varepsilon\right)$, the Green's function is trivial
\begin{equation}
G_{0,0}^{R}\left(  \varepsilon\right)  =\frac{1}{\varepsilon-\widetilde{E}_{0}
+{\mathrm{i}}(\,^{{L}}\Gamma_{0}+\,\,^{{R}}\Gamma
_{0}+\,\,^{{\phi}}\Gamma_{0})}. 
\label{eq_G-resonance}
\end{equation}
By taking the $\Gamma$'s independent on $\varepsilon$ in the range of
interest, we get the ``broad-band'' limit. We drop unneeded indices
and arguments for the time being. From this Green's function all the
transmission coefficients can be evaluated at the Fermi energy
\begin{eqnarray}
T_{{R,L}}  &  = & 4\,^{{R}}\Gamma\,\,\left|  G_{0,0}\right|
^{2}\,\,^{{L}}\Gamma,\label{eq_T-resonance}\\
T_{\phi,{L}}  &  = & 4\,\,^{{\phi}}\Gamma\,\left|  G_{0,0}\right|
^{2}\,\,^{{L}}\Gamma,\,\nonumber\\
T_{{R},\phi}  &  = & 4\,\,^{{R}}\Gamma\,\,\left|  G_{0,0}\right|
^{2}\,\,^{{\phi}}\Gamma.\nonumber
\end{eqnarray}
From the energy dependent transmittance we obtain the total transmittance
\begin{eqnarray}
T_{{R,L}}(\varepsilon)& = &4\,\,\,^{{R}}\Gamma\frac{1}{\left(
\varepsilon-\widetilde{E}_{0}\right)  ^{2}+(\,^{{L}}\Gamma+\,\,^{{R}}%
\Gamma+\,\,^{{\phi}}\Gamma)^{2}}\,\nonumber\\
&&\times  ^{{L}}\Gamma\,\left\{
1-\frac{^{{\phi}}\Gamma}{^{{L}}\Gamma+^{{R}}\Gamma
}\right\}  . \label{eq_Treson+DPmodel}%
\end{eqnarray}
The first term in the curly bracket is the coherent contribution while the
second is the incoherent one. We notice that the effect of the decoherence
processes is to lower the value of the resonance from its original one in a
factor
\begin{equation}
\frac{\left(  ^{{L}}\Gamma+\,^{{R}}\Gamma\right)  }%
{(^{{L}}\Gamma+\,\,^{{R}}\Gamma+\,\,^{{\phi}}\Gamma)}.
\label{eq_width-change}%
\end{equation}
In compensation, transmission at the resonance tails becomes increased.

It is interesting to note that if the resonant level lies between $\mu
_{o}+e{V}$ and $\mu_{o}$ and provided that the voltage drop
$e{V}$ is greater than the resonance width $(^{{L}}%
\Gamma+\,\,^{{R}}\Gamma+\,\,^{{\phi}}\Gamma),$ we take
$T_{{R,L}}(\varepsilon,e{V})\simeq T_{{R,L}}%
(\varepsilon)$ and we can easily compute the \emph{non-linear
response}$.$ Notably, one gets that the total current does not change
as compared with that in absence of decoherent processes, i.e.
\begin{eqnarray}
{\mathsf I}\frac{h}{2e}& = &\int_{\mu_{o}}^{\mu_{o}+e{V}}\widetilde
{T}_{{R,L}}(\varepsilon){d}\varepsilon\nonumber\\
& = &\int_{\mu_{o}}^{\mu
_{o}+e{V}}T_{{R,L}}^{o}(\varepsilon){d}\varepsilon\nonumber\\
& = &4\pi\,\,^{{R}}\Gamma\,\frac{1}{(^{{L}}\Gamma+\,\,^{{R}
}\Gamma)}\,^{{L}}\Gamma. 
\label{eq_non-linear-resonant-current}%
\end{eqnarray}
Thus, in this extreme quantum regime, the decoherence processes do not
affect the overall transport.

In spite of the complexity of the general problem of decoherence in
mesoscopic systems we extract an important lesson from the case of
resonant tunneling with decoherence solved above within the DP
model. The inclusion of external degrees of freedom has three effects

1) It broadens the resonance relaxing energy conservation.

2) The integrated intensity of the elastic (coherent) peak is decreased.

3) The inelastic contribution came out to compensate this loss and maintains
the value of the total transmittance integrated over energy.

4) Taking this system as representative of those whose spectra are
strongly quantized, i.e. with well defined, isolated resonances, one
may state that these systems are quite stable against decoherence.

\section{The solution of Time Dependence}

First we remember that the Green's function gives the possibility to
write non trivial initial values in the space-time coordinate
$X_{j}=({\mathbf r}_{j},t_{j})$ \ described by
$\varphi_{\mathrm{source}}(X_{j}).$ Therefore, any wave function
injected at space time $X_{j}=({\mathbf r}_{j},t_{j})$ by an arbitrary
source is propagated by the retarded Green's function
\begin{equation}
\varphi(X_{2})={\mathrm i}\hbar\int G^{R}(X_{2};X_{j})\varphi_{\mathrm{source}}%
(X_{j}){\mathrm d}X_{j} \label{eq_evolution1}%
\end{equation}
Here, instead of using a discrete spatial index, we adopted the notation
$G_{{\mathbf r}_{2},{\mathbf r}_{j}}^{R}(t_{2;}t_{j})\rightarrow G^{R}%
(X_{2};X_{j})$ and $\sum_{{\mathbf r}_{j}}\int{\mathrm d}t_{j}\rightarrow
\int{\mathrm d}X_{j}.$ To inject particles with some definition in momentum
and/or energy we need to introduce precise correlation between space and time
variables. We resort to the complementary equation
\begin{equation}
\varphi^{\ast}(X_{1})=-{\mathrm i}\hbar\int\varphi_{\mathrm{source}}^{\ast
}(X_{k})G^{A}(X_{k};X_{1})\mathrm{d}X_{k}, \label{eq_evolution2}%
\end{equation}
which we can use to find how the density matrix of particles depends on both
correlated initial conditions
\begin{eqnarray}
& &\left[  \varphi^{\ast}(X_{1})\varphi(X_{2})\right]  =\hbar^{2}\int\int
G^{R}(X_{2,}X_{j})\nonumber\\
& &\times \left[  \varphi_{\mathrm{source}}(X_{j})\varphi
_{\mathrm{source}}^{\ast}(X_{k})\right]  G^{A}(X_{k,}X_{1})\mathrm{d}%
X_{j}\mathrm{d}X_{k} \label{eq_densidad}%
\end{eqnarray}
The usual density function is defined by taking $X_{1}=X_{2}$ i.e.
$\rho\left(  X_{1}\right)  =\varphi^{\ast}(X_{1})\varphi(X_{1})$. This is
nothing else but the Schr\"{o}dinger equation written in a general form that allows arbitrary boundary conditions $\left[  \varphi_{\mathrm{source}}%
(X_{j})\varphi_{\mathrm{source}}^{\ast}(X_{k})\right]  $ . One is used to the
initial conditions at space-time coordinate $X_{i}$ \ of the form
$\rho_{\mathrm{source}}\left(  X_{i}\right)  =\left[  \varphi_{\mathrm{source}%
}(X_{i})\varphi_{\mathrm{source}}^{\ast}(X_{i})\right]  $ which allows
precision in the specification of position and time at the price of absolute
uncertainty in momentum and energy. The formalism allows one to program an
uncertainty trade-off \ to approach better to a semiclassical initial
condition with energy and momentum known up to some precision.

The basic idea to get the physics of time dependent phenomena from
Eqs.  (\ref{eq_densidad}) or from its generalization in Quantum Fields
Theory\cite{z--Keldysh,z--Danielewicz} is to recognize that in any
Green's function $G
(t_{j};t_{k})$, a macroscopically observable time is
$t=\frac{1}{2}\left[ t_{j}+t_{k}\right] $ whose Fourier transform is
an observable frequency $\omega.$\ Meanwhile, system's microscopic energies, $\varepsilon$, are Fourier transforms
of internal time differences $t_{j}-t_{k}$. 
By operating carefully with this concept in Eq. (\ref{eq_densidad}), the dynamical
transmittances $T({\mathbf r}
_{f},{\mathbf r}_{i};\varepsilon,\omega)\equiv
T_{f,i}(\varepsilon,\omega)$ with ${\mathbf r}_{i}$ a position in
channel $i$ and ${\mathbf r}_{f}$ a position in channel $f$ are
obtained. The manipulation of the time integrals is quite subtle and we refer to
Ref.\cite{z--GLBE2} for details. The basic result is
\begin{equation}
T_{f,i}(\varepsilon,\omega)=2\Gamma_{f}G_{f,i}^{R}(\varepsilon+\frac{1}%
{2}\hbar\omega)2\Gamma_{i}G_{i,f}^{A}(\varepsilon-\frac{1}{2}\hbar\omega),
\label{eq_T(w)}
\end{equation}
and is consistent with the steady-state Fisher-Lee equation.
From this new formula we can evaluate the probability that the wave packet of mean energy
$\varepsilon$ propagates from $i$ to $f$ in a time $t$ as
\begin{equation}
T_{f,i}(\varepsilon,t)=\int T_{f,i}(\varepsilon,\omega)\exp\left[
-\mathrm{i}\omega t\right]  \frac{\mathrm{d}\omega}{2\pi}. \label{eq_T(t)}%
\end{equation}
This time dependent transmittance was incorporated in a generalization
of the Landauer formula for time dependent phenomena.

Since the spectrum is continuous we can keep the lowest order in the
frequency expansion and obtain
\begin{equation}
T_{f,i}(\varepsilon,\omega)\simeq\frac{T_{f,i}(\varepsilon)}{1-\mathrm{i}%
\omega\tau_{P}}. \label{eq_T(w) approx}%
\end{equation}
Clearly, such approximation would give an exponential for the
propagation dynamics which does not describe the very short time
regime. According to Eq.(4.5) in Ref. \cite{z--GLBE2}, \ the typical
propagation time $\tau_{P}$ between points ${\mathrm r}_i$ and ${\mathrm r}_f$ results
\begin{eqnarray}
\tau_{P}  &  = &\frac{{\mathrm i}\hbar}{2}\left[  G_{f,i}^{R}(\varepsilon
)\frac{\partial}{\partial\varepsilon}G_{f,i}^{R}(\varepsilon)^{-1}
+G_{i,f}^{A}(\varepsilon)\frac{\partial}{\partial\varepsilon}G_{i,f}
^{A}(\varepsilon)^{-1}\right] \label{eq_tau_P}\\
&  = &-\frac{{\mathrm i}\hbar}{2}\frac{\partial}{\partial\varepsilon}\ln
\frac{G_{f,i}^{R}(\varepsilon)}{G_{i,f}^{A}(\varepsilon)}\nonumber
\end{eqnarray}
This was evaluated in various simple systems in
Ref.\cite{z--GLBE2}. There, for {\textit ballistic metals} with velocity $v_{\varepsilon}$ one gets 
$\tau_P=\left|{\mathbf r}_f-{\mathbf r}_i \right|/v_{\varepsilon}$. For {\textit diffusive metals} where impurity collision at a rate $1/\tau_o$ determines the diffusion constant $D_{\varepsilon}=v_{\varepsilon}^2\tau_o/2$
it results $\tau_P=\left|{\mathbf r}_f-{\mathbf r}_i \right|^2/(2D_{\varepsilon})$.

As a striking example, we mention the ``simple'' case of
\textit{tunneling
\ through a barrier}\cite{z--Landauer-time} of length $L$ and height $U$
exceeding the kinetic energy $\varepsilon$ \ of the particle. One gets
\begin{equation}
\tau_{P}=\frac{L}{\sqrt{\frac{2}{m}(U-\varepsilon)}}, \label{eq_tau-tunneling}%
\end{equation}
which, within our non-relativistic description, can be extremely short
(even superluminal\cite{z--time-superluminal}!) provided that the
barrier is high enough.

In a double barrier system, in the regime of \textit{resonant tunneling,} the
above contribution becomes negligible and the propagation time is then fully
determined by the life-time inside the well. In fact, by using the functions
of the previous Section, one gets a delay of
\begin{equation}
\tau_{P}=\frac{\hbar}{2(\,^{{L}}\Gamma+\,^{{R}}\Gamma)}
\label{eq_tau_resonance}%
\end{equation}
which readily limits the admittance associated to the device to ${\mathsf G}
{(\omega)}={\mathsf G}{(0)}/(1-{\mathrm i}\omega\tau_{P})$. This is in fair
agreement with the experimental results\cite{z--time/exp:Sollner}. Tunneling
times can also be calculated in more complex situations such as disordered
\cite{z--time-chains}systems and situations with coherent capacitive effects
and phonon interactions\cite{z--Jauho-book} and
decoherent\cite{z--GLBE2}processes..

\section{Final Remarks.}

We hope these lectures have been able to convey, at least partially,
the essence of our message: In our attempt to understand the quantum
world it is possible, and convenient, to make very simple models of
nature. Those models, though not necessarily complete, can show us
many effects that surprise our classical intuition and which would
have been obscured by other more ``complete'' descriptions. Every one
of these surprises can give a new twist in the experimental research,
which is indeed the ultimate truth, and eventually become the source of an
innovative application.
        
In our toolbox to extract information from the proposed models the most
important elements have been the decimation method and the Green's
functions technique. Both are intimately connected with the
renormalization group concept. As such they are fundamental in
generating the models themselves, as they help us to identify the
relevant variables.  Simultaneously, our understanding of the quantum
world has benefited from fresh approaches to long standing problems as
that of the tunneling time. In the process of computing transport
properties we have also learnt the deepest principles of statistical
mechanics as to what is the origin of macroscopic irreversibility and
established the conditions for the validity of the ergodic
hypothesis. The fundamentals of decoherent processes in transport and
their consequences are currently of intense interest. The lines of
thought we have sketched here, have turned out to be very productive in
this direction, and we hope they will continue being fruitful through the
action of our students.

\end{multicols}

\end{document}